\documentclass[aps,pra,twocolumn,amsmath,amssymb,nofootinbib,superscriptaddress]{revtex4-2}
\usepackage{times}
\usepackage[pdftex]{graphicx}
\usepackage{dcolumn}
\usepackage{bm}
\usepackage{amsmath}
\usepackage{braket}
\usepackage{indentfirst}
\usepackage{float}
\usepackage[colorlinks]{hyperref}
\usepackage[dvipsnames]{xcolor}
\usepackage{xcolor}

\renewcommand\vec{\mathbf}
\definecolor{aqua}{RGB}{69,139,116}
\definecolor{lgreen}{RGB}{0,102,0}

\newcommand{\rev}[1]{\textcolor{black}{#1}}

\usepackage{subfigure}
\usepackage{algorithm}
\usepackage{algorithmicx}
\usepackage{algpseudocode}
\usepackage{amsmath}
\usepackage{verbatim}
\usepackage{makecell}
\usepackage[normalem]{ulem}
\begin{document}

\title{Evaluating the Resilience of Variational Quantum Algorithms to Leakage Noise}

\author{Chen Ding}
\affiliation{Henan Key Laboratory of Quantum Information and Cryptography, Zhengzhou, Henan 450000, China}
\author{Xiao-Yue Xu}
\author{Shuo Zhang}
\affiliation{Henan Key Laboratory of Quantum Information and Cryptography, Zhengzhou, Henan 450000, China}
\author{He-Liang Huang}
\email{quanhhl@ustc.edu.cn}
\affiliation{Henan Key Laboratory of Quantum Information and Cryptography, Zhengzhou, Henan 450000, China}
\affiliation{Hefei National Research Center for Physical Sciences at the Microscale and School of Physical Sciences, University of Science and Technology of China, Hefei 230026, China}
\affiliation{Shanghai Research Center for Quantum Science and CAS Center for Excellence in Quantum Information and Quantum Physics, University of Science and Technology of China, Shanghai 201315, China}
\affiliation{Hefei National Laboratory, University of Science and Technology of China, Hefei 230088, China}
\author{Wan-Su Bao}
\email{bws@qiclab.cn}
\affiliation{Henan Key Laboratory of Quantum Information and Cryptography, Zhengzhou, Henan 450000, China}



\begin{abstract}
As we are entering the era of constructing practical quantum computers, suppressing the inevitable noise to accomplish reliable computational tasks will be the primary goal. Leakage noise, as the amplitude population leaking outside the qubit subspace, is a particularly damaging source of error that error correction approaches cannot handle. However, the impact of this noise on the performance of variational quantum algorithms (VQAs), a type of near-term quantum algorithms that is naturally resistant to a variety of noises, is yet unknown. Here, {we consider a typical scenario with the widely used hardware-efficient ansatz and the emergence of leakage in two-qubit gates}, observing that leakage noise generally reduces the expressive power of VQAs. Furthermore, we benchmark the influence of leakage noise on VQAs in real-world learning tasks. Results show that, both for data fitting and data classification, leakage noise generally has a negative impact on the training process and final outcomes. Our findings give strong evidence that VQAs are vulnerable to leakage noise in most cases, implying that leakage noise must be effectively suppressed in order to achieve practical quantum computing applications, whether for near-term quantum algorithms and long-term error-correcting quantum computing.

\end{abstract}

\maketitle

\section{Introduction}

While quantum computing is one of the most promising new trends in information processing, the noisy qubits in the \textit{Noisy Intermediate-Scale Quantum} (NISQ) era~\cite{superconducting_2020review, google_supremacy,ustc_supremacy,wu_strong_2021,zhu2022quantum,Verifying_RQC} limit its immediate applicability. Thus, correcting, mitigating, or avoiding the quantum errors to achieve valid computation results is a major topic in current era~\cite{fowler2012surface, chen_exponential_2021,zhao_realizing_2021,erhard2021entangling,andersen_repeated_2020,marques2021logical, krinner2021realizing, PhysRevLett.127.090502,9code,huang2021emulating,unfolding_readout_noise,Mitigating_readout_noise,nation_scalable_2021,smith_qubit_2021,ding2021noiseresistant}. Variational quantum algorithms (VQAs)~\cite{peruzzo_variational_2014,experimental_QGAN,qccnn,Schuld2019QuantumML,biamonte2017quantum,havlivcek2019supervised,saggio2021experimental,cerezo_variational_2021,PhysRevX.10.021067} have emerged as the one of leading candidates to achieve application-oriented quantum computational advantage on NISQ devices, owing to their hybrid quantum-classical approach which has potential noise resilience~\cite{mcclean_theory_2016,1gentini_noise-resilient_2020,3sharma_noise_2020}. To date,  VQAs have already been demonstrated successfully for a variety of machine learning tasks, such as classification~\cite{cong2019quantum,havlivcek2019supervised,qccnn}, generative modeling~\cite{lloyd2018quantum, experimental_QGAN}, optimization~\cite{harrigan2021quantum}, and quantum chemistry problems~\cite{kandala_hardware-efficient_2017,google2020hartree,Romero_2018,PhysRevB.102.235122,PhysRevX.8.031022,nam_ground-state_2020}. 

Despite the continuing improvement of such experimental demonstrations, a critical question remains as to how noise affects the performance of VQAs. This issue has been explored extensively in recent years, with many valuable findings. Generally, the resillience of VQAs exists for a wide class of noise, as analyzed in Refs.~\cite{mcclean_theory_2016,1gentini_noise-resilient_2020,3sharma_noise_2020}. Yet such noise resillience is highly limited as the noise level rising. As suggested in~\cite{5saib_effect_2021,7wright_numerical_2021,8gowrishankar_numerical_2021,9wang_noise-induced_2021,6ito_universal_nodate}, large noise may leads to problems such as performance degradation or barren plateaus.
Among all the works mentioned above, leakage error is not included in their error models, which is a notable omission given the severity of leakage error. Leakage error refers to the accidental qubit activations outside of the computational space in two-level quantum computers. Usual quantum error correction protocols cannot correct (and may even exacerbate) leakage error~\cite{9259963,ghosh_understanding_2013}. Even if the probability of these leakage errors is modest, they will accumulate and eventually corrupt the error correction code. Thus, analyzing the impact of leakage error on VQAs is crucial for the near-term application of quantum computing.

In this paper, we investigate performance of VQAs with the occurrence of leakage error. We first analyze the effect model of leakage error in the execution of quantum circuits, and propose an expressibility measure that predicts the performance of variational ansatz with noiseless environment and leakage error. {Next, based on the assumption of the leakage occurrence during two-qubit gates, the most typical leakage noise in superconducting quantum computing, we conduct numerical experiments to benchmark the impact of leakage error on applications of data fitting and data classification when using the widely used hardware-efficient ansatz,} with varying the system size, circuit depth, and leakage probability. The simulation results of expressibility, data fitting and data classification all suggest VQAs are vulnerable to leakage error. Moreover, our results on chain, ladder and lattice processor also show that better connectivity architectures also do not contribute to leakage error immunity. This work facilitates a deeper understanding of how noise impacts the performance of VQAs, and provides new insight on the key issues of NISQ techniques.

\section{Theoretical Analysis}

In this section, we will theoretically analyze the impact of leakage error on VQAs, including the measurement results and the expressibility.

\subsection{Leakage Error in Quantum Circuits}\label{subsec:leakage_error_in_VQA}

As stated above, the phenomenon of leakage refers to the state activation out of the usual two-level computation space $\ket{0}$ and $\ket{1}$.
Let's first take a look at how leakage error affects the output of quantum computing.
Take one-qubit for instance, the leakage noise introduces activated states as
\begin{equation*}\label{eqn:state}
\ket{\psi}=\alpha_0\ket{0}+\alpha_1\ket{1}+\alpha_2\ket{2}+\alpha_3\ket{3}+...
\end{equation*}

If high energy levels of the qubits, $\ket{2}$, $\ket{3}$, etc, can be readout properly, one can directly utilize the post-selection strategy to rule out events with leakage errors. Thus, here we assume that states leaking to higher energy levels cannot be readout correctly, where their probabilities of being misread as $\ket{0}$ and $\ket{1}$ are $\beta$ and $1-\beta$, respectively. {In this case, the read populations of $\ket{0}$ and $\ket{1}$ are
\begin{align*}
    p_0&=|\alpha_0|^2+\beta(1-|\alpha_0|^2-|\alpha_1|^2),\\
    p_1&=|\alpha_1|^2+(1-\beta)(1-|\alpha_0|^2-|\alpha_1|^2).
\end{align*}}

As a consequence, the expectation of any observable in $\ket{\psi}$ is deviated from its original value. For instance, the expectation of $O=\text{Diag}(\lambda_0,\lambda_1)$ will be
\begin{align*}
&\braket{\psi|O|\psi}\\
=&\lambda_0(|\alpha_0|^2+\beta(1-|\alpha_0|^2-|\alpha_1|^2))\\
+&\lambda_1(|\alpha_1|^2+(1-\beta)(1-|\alpha_0|^2-|\alpha_1|^2))\\
=&(|\alpha_0|^2+|\alpha_1|^2)\braket{\phi|O|\phi}\\
+&(1-|\alpha_0|^2-|\alpha_1|^2)(\lambda_0\beta+\lambda_1(1-\beta)),\label{eqn:biased_expectation}
\end{align*}
in which $\ket{\phi}=\frac{1}{\sqrt{|\alpha_0|^2+|\alpha_1|^2}}(\alpha_0\ket{0}+\alpha_1\ket{1})$.

For non-diagonal observable $H$, there are two ways to perform the measurement---diagonalization and Pauli decomposition. For the diagonalization method, we implement the measurement by rotating the measurement basis to diagonalize $H$:
\[VHV^\dagger=\begin{pmatrix}
    \lambda_0 & 0\\
    0 & \lambda_1
    \end{pmatrix}\]
the expectation will be
\begin{align*}
&\braket{\psi|H|\psi}\\
=&\lambda_0(|V_{00}\alpha_0+V_{01}\alpha_1|^2+\beta(1-|\alpha_0|^2-|\alpha_1|^2))\\+&\lambda_1(|V_{10}\alpha_0+V_{11}\alpha_1|^2+(1-\beta)(1-|\alpha_0|^2-|\alpha_1|^2))\\
=&(|\alpha_0|^2+|\alpha_1|^2)\braket{\phi|H|\phi}\\
+&(1-|\alpha_0|^2-|\alpha_1|^2)(\lambda_0\beta+\lambda_1(1-\beta)).
\end{align*}

For the Pauli decomposition method, we implement the measurement by sum of Pauli matrices $H=\sum h_i\sigma_i$, the expectation will be
\begin{align*}
&\braket{\psi|H|\psi}\\
=&h_0+h_1\braket{\psi|\sigma_x|\psi}+h_2\braket{\psi|\sigma_y|\psi}+h_3\braket{\psi|\sigma_z|\psi}\\
=&(|\alpha_0|^2+|\alpha_1|^2)\braket{\phi|H|\phi}\\
+&(1-|\alpha_0|^2-|\alpha_1|^2)(h_0+(h_1+h_2+h_3)(2\beta-1)).
\end{align*}
We observe the two implementation methods will generally yield different expectation value for the same non-diagonal observable, which is due to the misreading mechanism stated above, and the limitation of rotation $V$ in the 2-level computational space. Such results suggest the leakage state after a misreading channel cannot simply be viewed as pure or mixed 2-level states.

The direct effect of leakage is the deviation of the measurement results, determined by the population of higher-energy states, the misreading probability $\beta$ and $1-\beta$, the observable $H$, the measurement implementation. 
This gives us an intuitive impression on the impact of leakage error on quantum computing. As more qubits and gate operations are introduced, the analysis will become complicated, especially for VQAs, which would adjust the parameters of quantum operations during the training procedure. We thus propose a more powerful tool to evaluate the performance of VQAs with leakage error, by estimating the expressibility.

\subsection{The Expressibility Measure for VQAs with Leakage error}\label{subsec:expressibility}

\begin{figure}[t]
    \begin{center}
    \includegraphics[width=0.8\linewidth]{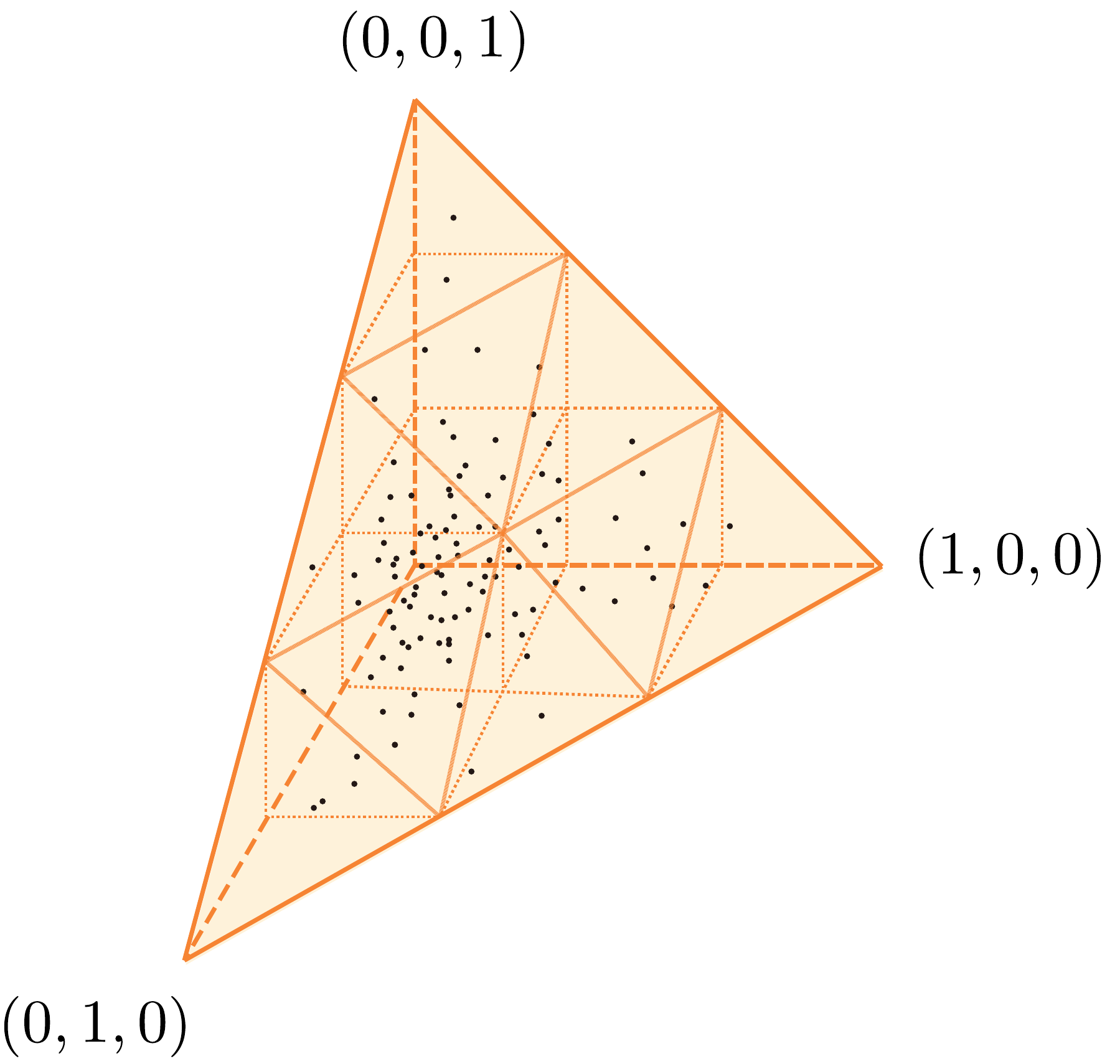}
    \end{center}
    \caption{\textbf{The 3-dimensional simplex $\Delta^{4}$ and uniform distribution samples in it (corresponding to system size $n=2$).} The black dots are the samples $\vec{F}_\text{uniform}$, which distributes evenly inside the simplex. A well-performing ansatz should output similar samples. Our expressibility measure evaluate minus MMD distance between $\vec{F}_\text{uniform}$ and the ansatz's output samples. The dashed lines inside the simplex are helpers to view the shape in depth.}
    \label{fig:cubes}
    \end{figure}

\begin{figure}[t]
    \begin{center}
    \includegraphics[width=1\linewidth]{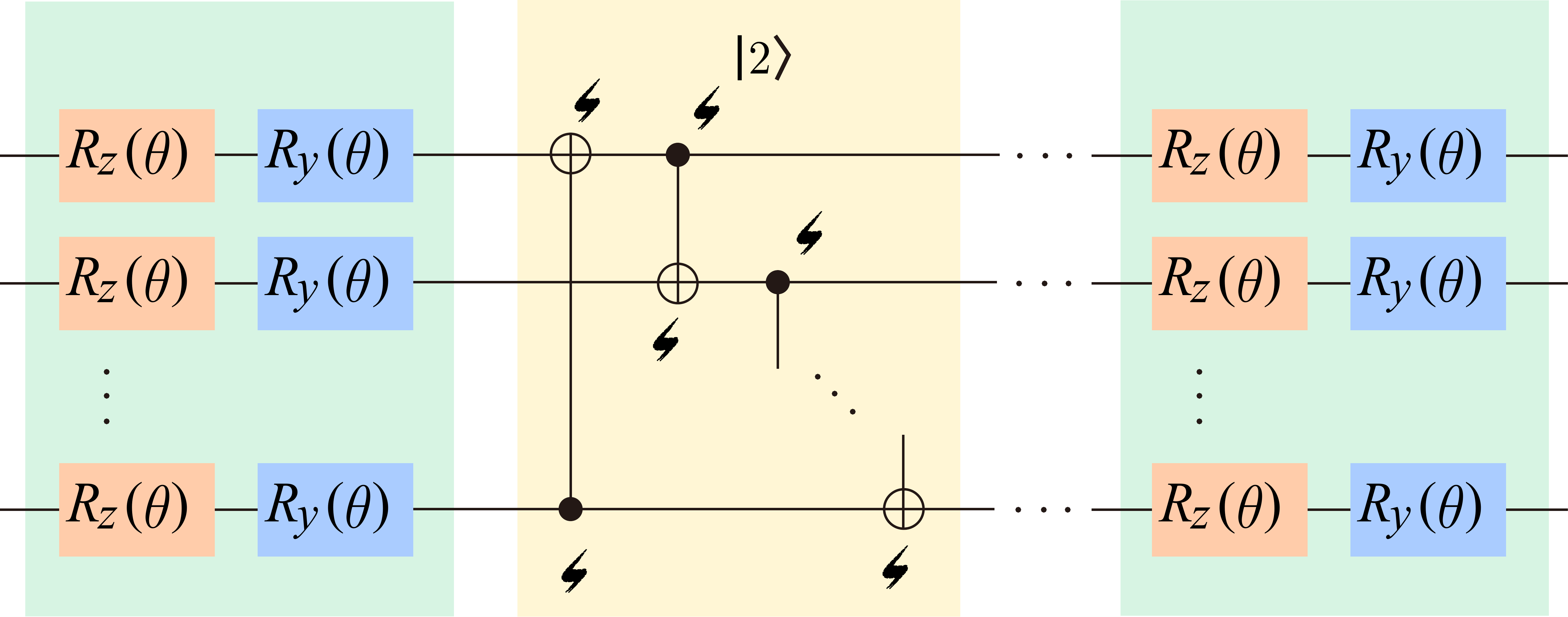}
    \end{center}
    \caption{\textbf{A multi-layer ansatz for VQA}, in which we only consider leakage error for the two-qubit gates in the entangling layer (shown in the light yellow area). In the simulation, we assume the CNOT gates are implemented by a CZ gate and two Hadamard gates, as $\text{CNOT}=(I\otimes H)\text{CZ}(I\otimes H)$.}
    \label{fig:vqc}
    \end{figure}

VQAs utilize parameterized quantum circuit (PQC) models that can be trained for a variety of machine learning tasks. \rev{Given a fixed circuit structure, the operation of the PQC is parameterized by the variables $\vec{\theta}$, as illustrated in Fig.~\ref{fig:vqc}. We therefore denote the map from the parameters $\vec{\theta}$ to the quantum circuit as $U(\cdot)$. To predict the performance of the corresponding VQAs before applying the PQC to a certain application, we need to evaluate the expressibility (or expressivity) of the map $U(\cdot)$~\cite{PRXQuantum.3.010313,5saib_effect_2021,tangpanitanon_expressibility_2020,nakaji_expressibility_2021,ssim_expressibility}, which indicates its ability to generate eligible outputs for a general purpose.} For instance, Sukin Sim $et~al$. quantify the expressibility by estimating the statistical properties of a PQC's outputs, given uniformly distributed parameters $\theta$ and a fixed initial state~{\cite{ssim_expressibility}}. Specifically, they calculate the Kullback-Leibler (KL) divergence~\cite{10.1214/aoms/1177729694} of two distribution functions
\begin{equation}\label{exp1}
    \text{Expr}_1(U)=D_{\text{KL}}\left(f_{F_U}||f_{F_\text{Haar}}\right).
\end{equation}

In Equation~(\ref{exp1}), the PQC is assigned with randomly generated parameters sampled from uniform distribution in $[0,2\pi]$. $f_{F_U}$ is the distribution function of the overlap $F_U=|\braket{\psi_1|\psi_2}|^2$ of the PQC's arbitrary two outcomes given random parameters $\theta$, and $f_{F_\text{Haar}}$ is the distribution function of the overlap $F_\text{Haar}=|\braket{\phi_1|\phi_2}|^2$ of arbitrary two states in the uniform distribution (the Haar distribution) of the Hilbert space. The initial state of the ansatz is fixed as $\ket{\vec{0}}$. By comparing the fidelity distribution $F_U$ and $F_\text{Haar}$, Expr$_1$ implies the ansatz's capability of approximately generating outputs in Haar distribution, which meet the demand of applications like finding the ground state for an arbitrary Hamiltionian, or preparation of an arbitrary pure state.

However, Expr$_1$ assumes the ansatz's outputs are pure states, \rev{which is no more suitable for expressibility evaluation in the leakage case. As shown in Subsection~\ref{subsec:leakage_error_in_VQA}, the expectation value of misread leakage state varies with the choice of measurement basis.
} To remove the Hilbert output space assumption of Expr$_1$, we consider the outputs as real-valued vectors---the population of basis states. By computational basis measurement, the higher-energy states are read in a fixed way, which can then be viewed as the natural output of the ansatz. Any state with leakage solely corresponds to one real vector in the restricted hyperplane in $\mathbb{R}^{2^n}$, which is also a standard simplex in $\mathbb{R}^{2^n-1}$:
\begin{align}
\Delta^{2^n}:=&\{(p_0,p_1,...,p_{2^n-2})\in\mathbb{R}^{2^n-1}|\sum^{2^n-2}_{i=0}p_i<1,\nonumber \\
&p_i\geq 0\quad \forall i=0,...,2^n-2\}.
\end{align}

We thus propose a real-valued output version method for expressibility estimation, as the uniformity of the ansatz outputs in the simplex $\Delta^{2^n}$. {Moreover, to cover higher-order statistics of the ansatz's output distribution $\vec{F}_U$, we directly calculate the distance between $\vec{F}_U$ and the uniform distribution $\vec{F}_\text{uniform}$ in the simplex. For achieving efficient and robust high-dimensional distance evaluation, we choose the Maximum Mean Discrepancy (MMD)~\cite{gretton_kernel_2012,DBLP:journals/corr/abs-2106-14277}
\begin{equation}
    \text{MMD}(P,Q):=\|\mathbb{E}_{X\sim P}[\varphi(X)]-\mathbb{E}_{\rev{Y\sim Q}}[\varphi(Y)]\|_{\mathcal{H}},
\end{equation}
as the distance measure, in which $\varphi(\cdot)$ is the function that maps \rev{the} samples $X,Y$ to the reproducing kernel Hilbert space. The expressibility of the ansatz $U(\vec{\theta})$ is defined as the minus MMD distance, as
\begin{equation}
    \text{Expr}_2(U)=-\text{MMD}(\vec{F}_U,\vec{F}_\text{uniform}).
\end{equation}
By definition, it is easy to find $\text{Expr}_2=0$ if and only if $\vec{F}_U=\vec{F}_\text{uniform}$, while a similar proposition is yet to be proven for $\text{Expr}_1$~\cite{ssim_expressibility}. Replacing $\varphi(\cdot)$ by a kernel $k(\vec{x},\vec{y})=\braket{\varphi(x),\varphi(y)}_{\mathcal{H}}$, it is then efficient to estimate the MMD with the samples of $\vec{F}_U$ and $\vec{F}_\text{uniform}$ (noted as $X_i,Y_i,i=1,...,N$, respectively), as 
\begin{align}
    &\text{MMD}(P,Q)\nonumber\\
    \approx&\frac{1}{N^2}\left|\sum_{i=1,j=1}^{N} k(X_i,X_j)+k(Y_i,Y_j)-2k(X_i,Y_j)\right|,
\end{align}
in which we use the kernel $k(\vec{x},\vec{y})=e^{-\frac{\|\vec{x}-\vec{y}\|^2}{4\sigma}}$.} The parameter $\sigma$ is taken as $0.01$ in our calculation to optimize the prediction performance. The initial state of the ansatz is taken as $\ket{\vec{+}}$. 
\rev{Our proposed method enables us to comprehensively evaluate whether the output of the variational quantum circuit with different circuit parameters can be well spread into the entire output space (the simplex), that is, the high expressibility.} 
\rev{Since we assume the outputs are real vectors, Expr$_2$ is more suitable for predicting ansatz's performance in machine learning applications like real data fitting or classification.}

\section{Numerical Experiments}

\begin{figure*}[htbp!]
\begin{center}
\includegraphics[width=\linewidth]{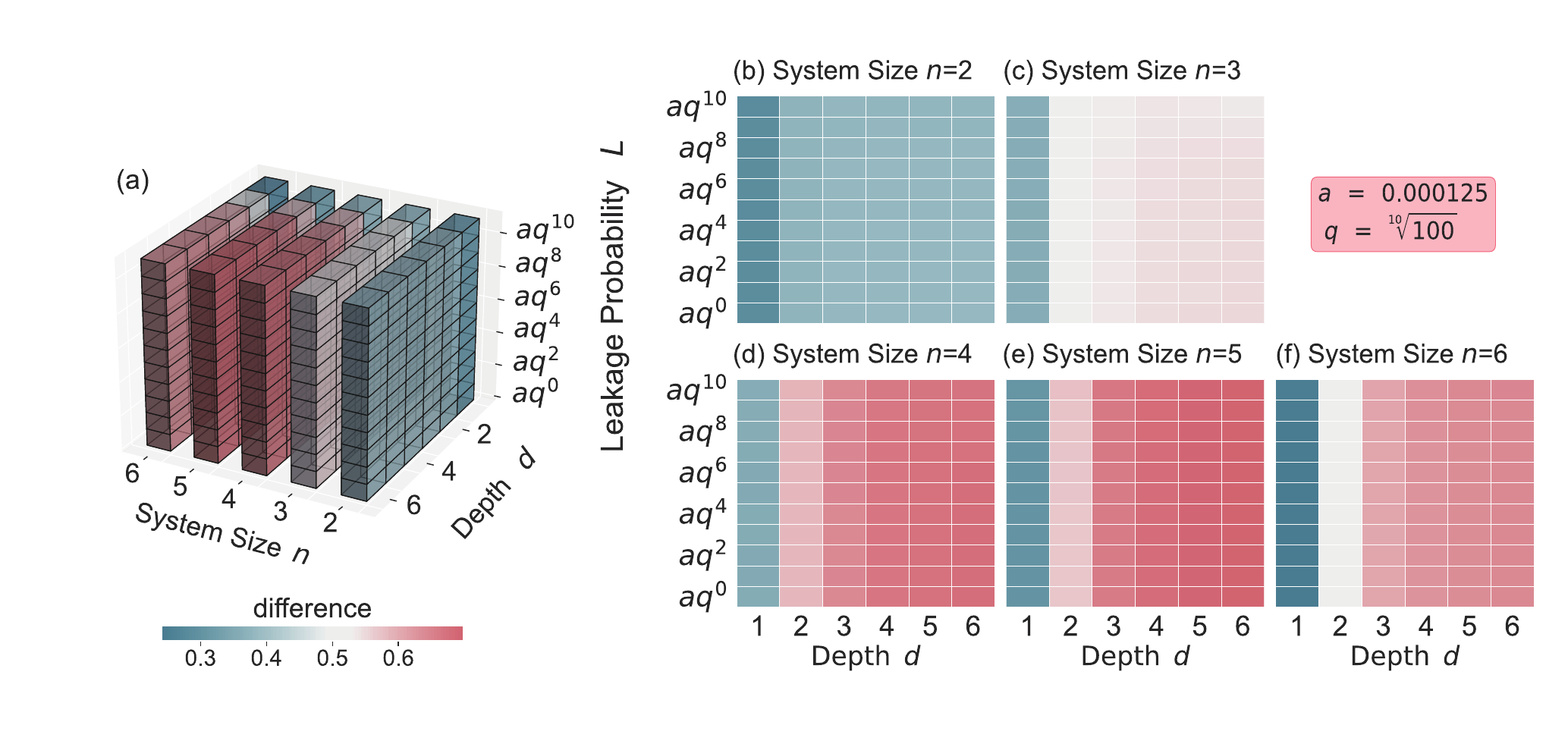}
\end{center}
\caption{\textbf{The expressibility difference between ideal ansatz and ansatz with leakage error, varying with system size $n$, circuit depth $d$, and leakage probability $L$.} In (a), we show a overall 3D heatmap of the expressibility difference. In (b-f), the five subfigures are the slices taken from (a), showing the case of different system sizes. In these figures, each lattice shows the average difference over 20 independent experiments. The average standard deviation of the means is about $5\cdot 10^{-4}$.}
\label{fig:expressibility}
\end{figure*}
The performance of VQAs heavily depends on the expressibility of the employed ansatz. However, different machine learning tasks and training methods may have different requirements for the output quantum states. We thus conduct numerical experiments to benchmark the actual impact of leakage error in various machine learning tasks, and validate the capability of prediction by our expressibility measure.
{For the sake of typicality, we consider the most commonly used hardware-efficient PQC ansatz~\cite{zoufal_quantum_2019,PhysRevA.103.032607,PhysRevA.101.032308,PhysRevA.103.032430,RevModPhys.94.015004}, shown in Fig.~\ref{fig:vqc}}, which consists of alternating trainable single-qubit layers and fixed entangling layers. For superconducting transmons, two-qubit gates constitute the dominant source of leakage, while single-qubit gates and measurement operations have negligible leakage probabilities which we ignore in the simulation. We model the leakage in CZ gates as an exchange between $\ket{11}$ and $\ket{02}$ during the CZ gate, i.e., $|11\rangle  \mapsto \sqrt {1 - 4L} |11\rangle  + {e^{i\phi }}\sqrt {4L} |02\rangle $ and $|02\rangle  \mapsto \sqrt {1 - 4L} |02\rangle  - {e^{ - i\phi }}\sqrt {4L} |11\rangle$, in which leakage probability $L$ is typically $1.25\cdot 10^{-3}$~\cite{varbanov_leakage_2020,PhysRevA.97.032306}. Besides, we also assume the transitions phase $\phi=0$ for simplicity. \rev{The misread coefficient $\beta$ is experimentally tunable and may be set as some specific values to minimize the effects of leakage in a certain computing task. We focus on the simplest case where higher-energy states are misread as $\ket{1}$ ($\beta=0$) in the experiments, since it is generally more probable that higher-level state being misread as $\ket{1}$ other than $\ket{0}$ in most superconducting quantum systems.}
Eventually, there are only three hyperparameters in our experiments: the system size $n$, the depth of the circuit $d$, and the leakage probability $L$. In all the experiments below, we take uniformly random initial parameters at the start of training.

\subsection{Expressibility Estimation}\label{subsec:expressibility_est}

We first estimate the expressibility of the employed ansatz with noiseless environment and leakage error, respectively, by using the method proposed in Sec.~\ref{subsec:expressibility} In our simulation, the hyperparameters are chosen as: system size $n=2,...,6$, circuit depth $d=1,...,6$ (We count the depth as the number of layers, which consist of trainable single-qubit layer and a fixed entangling layer, in the circuit, as shown in Fig.~\ref{fig:vqc}.), leakage probability $L$ proportionally ranges from $1.25\cdot 10^{-4}$ to $1.25\cdot 10^{-2}$. Under each combination of hyperparameters, we take 10000 samples from the circuit's output distribution and the uniform distribution to estimate the expressibility. Figure~\ref{fig:expressibility} shows the expressibility difference between noiseless ansatz and ansatz with leakage error, $\text{Expr}_2^{\text{noiseless}}-\text{Expr}_2^{\text{leakage}}$. It can be observed that the noiseless ansatz always has a higher expressibility. Moreover, the expressibility difference rises as the leakage probability $L$ and circuit depth $d$ increase.

\subsection{Data Fitting}

\begin{figure*}[htbp!]
    \begin{center}
    \includegraphics[width=1\linewidth]{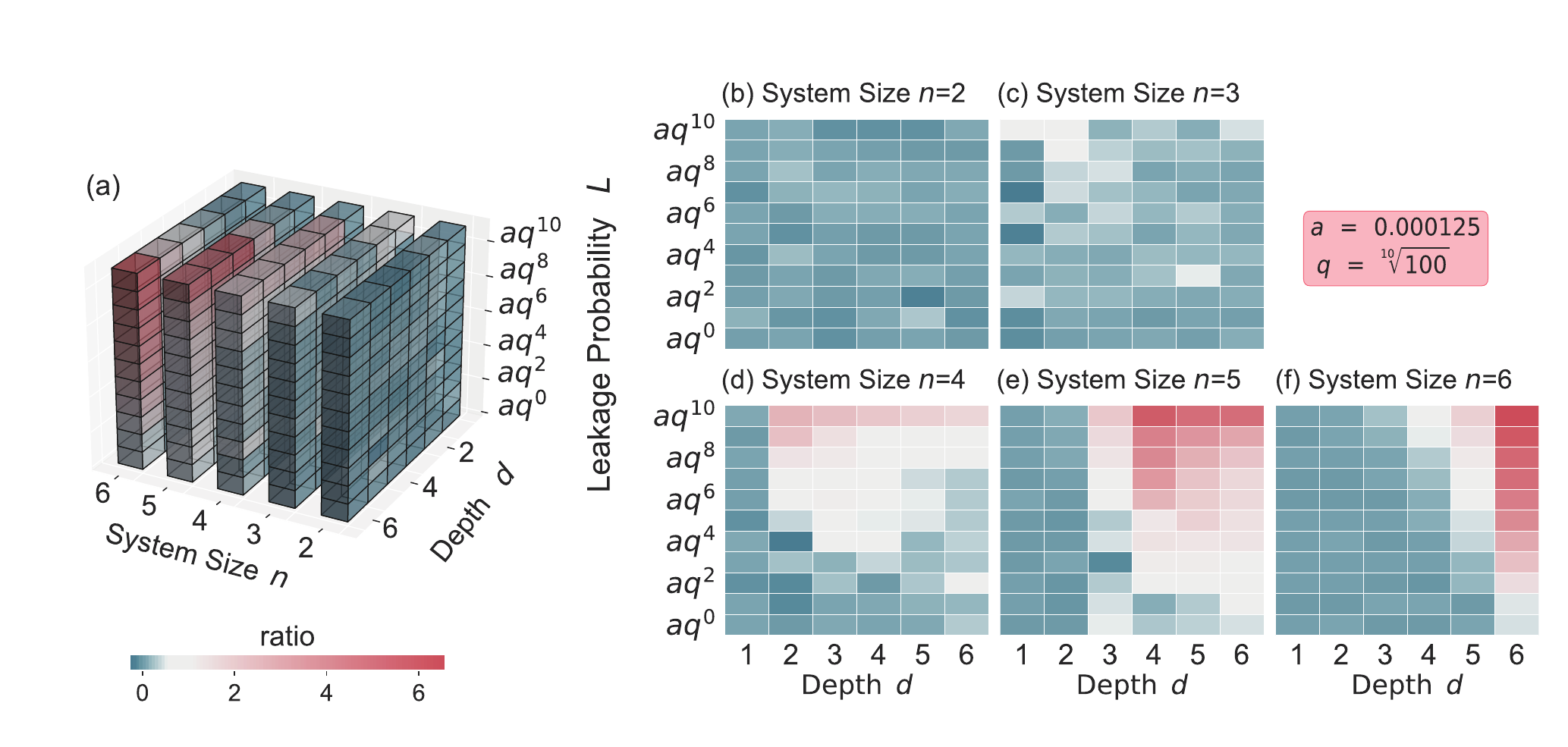}
    \end{center}
    \caption{\textbf{The logarithm of accuracy ratio of ideal ansatz and ansatz with leakage error, varying with system size $n$, circuit depth $d$, and leakage probability $L$.} In (a), we show a overall 3D heatmap of the ratio logarithm. In (b-f), the five subfigures are the slices taken from (a), showing the case of different system sizes. In these figures, each square shows the average result over 100 independent experiments.  The average standard deviation of the means is about $0.05$.}
    \label{fig:datafit}
    \end{figure*}

Next we compare the performance of VQAs with noiseless environment and leakage error on the application of data fitting---the most fundamental functionality of machine learning models. We set initial states as $\ket{+}^{\otimes n}$, and randomly sample vectors from the uniform distribution in $\Delta^{2^n-1}$ as the targets. The loss function is the fidelity between the target distribution $ \boldsymbol q$ and the output $\boldsymbol{p}(\boldsymbol{\theta})$ of VQA, as
\[f(\boldsymbol{\theta})=1-\left(\sum\sqrt{p_i(\boldsymbol{\theta})q_i}\right)^2.\]
where $\boldsymbol{\theta}$ is the trainable parameters in the PQC.

\begin{figure}[t]
\begin{center}
\includegraphics[width=1\linewidth]{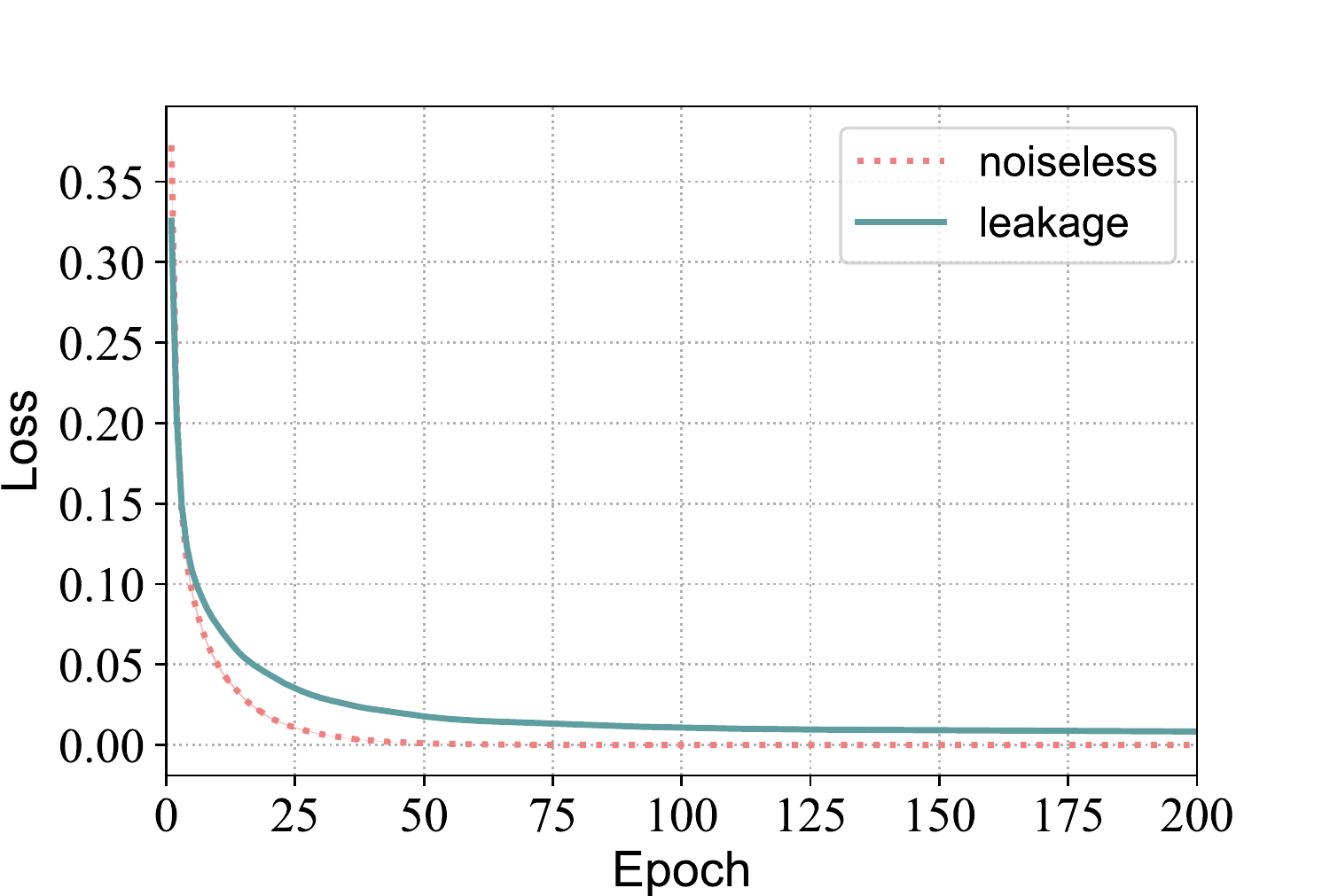}
\end{center}
\caption{\textbf{The training curve in the case of $n=d=6$, $L=0.0125$.} Each point (from Epoch 1 to Epoch 200) on the curve is an average result over 100 independent experiments. The standard deviations of the means for these independent experiment results are no more than $0.003$.  
}
\label{fig:onedatafit}
\end{figure}

Figure~\ref{fig:datafit} shows the logarithm (in base 10) of the ratio of the loss values of VQAs with noiseless environment and leakage error at epoch 200. The ratio rises as the leakage probability $L$ and circuit depth $d$ increase, suggesting that more leakage occurs with increasing depth and leakage probability. The highest ratio $10^{6.57}$ is achieved in the case of $n=d=6$, $L=0.0125$, while the lowest ratio $10^{-0.28}$ is at in the case of $n=4$, $d=2$, $L=0.000789$. In all the cases of these hyperparameters, the training of VQAs in the both noiseless and noisy environment converges. Figure~\ref{fig:onedatafit} shows the loss curve during the training in the case that highest ratio appears, which demonstrate the convergence. These results suggest that leakage error can degrade the performance of data fitting, this overall phenomenon is consistent with the prediction by expressibility.


\subsection{Data Classification on the Iris Dataset}

Furthermore, we investigate the effect of leakage error on a real-world application of VQA---the classification of Iris~\cite{10.2307/2394164,https://doi.org/10.1111/j.1469-1809.1936.tb02137.x}, which consists of 150 4-element vectors of three classes. We choose the \textit{virginica} and \textit{versicolor} classes for binary classification. The data vectors are encoded into amplitudes of 2-qubit quantum states as the circuit input. And we use the mean square error (MSE) cost function
\[C(\boldsymbol{\theta})=\sum_{i=1}^M(\braket{\psi_i|U(\boldsymbol{\theta})^\dagger Z^{\otimes 2}U(\boldsymbol{\theta})|\psi_i}-y_i)^2,\]
for training, in which $\boldsymbol{\theta}$ are the trainable parameters, $M=100$ is the number of samples, $\ket{\psi_i}$ are the encoded input quantum states and $y_i$ are the corresponding labels.

\begin{figure}[t]
\begin{center}
\includegraphics[width=1\linewidth]{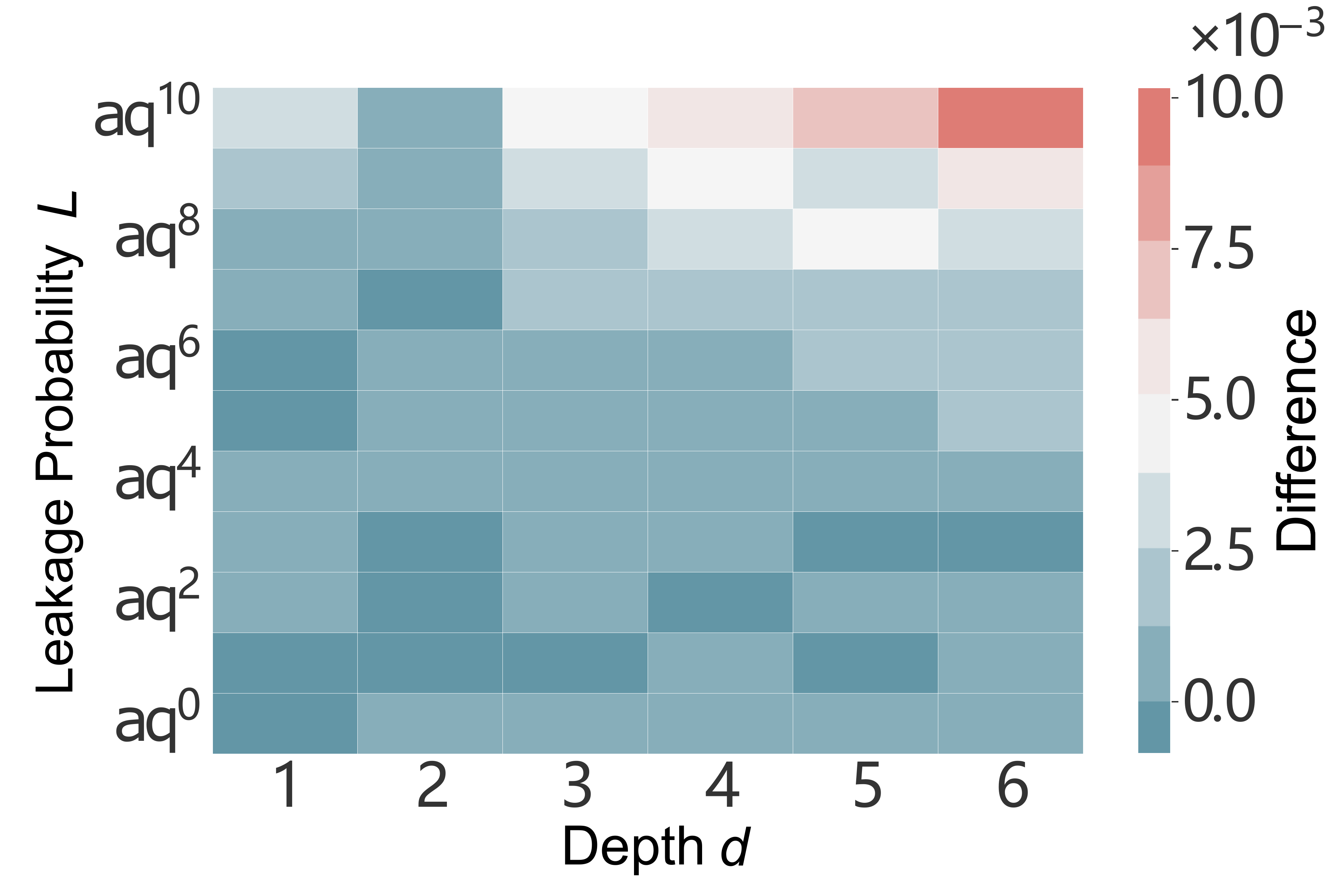}
\end{center}
\caption{\textbf{The cross validation score difference between ideal ansatz and ansatz with leakage error, with {varying} circuit depth $d$ and leakage probability $L$.} The system size $n$ is fixed as 2. Each lattice an average result over 100 independent experiments. The average standard deviation of the means is about $6\cdot 10^{-4}$.}
\label{fig:validation}
\end{figure}

We use 4-fold cross-validations for evaluating the performance of VQAs with noiseless environment and leakage error. Figure~\ref{fig:validation} shows the cross validation score difference between noiseless VQAs and VQAs with leakage error, at different circuit depth $d$ and leakage probability $L$. The difference rises as the leakage probability $L$ and circuit depth $d$ increase, also indicating that leakage error will degrade the performance of the classification, in consistency with the results in data fitting experiments.

\subsection{The Impact of Leakage Error on VQAs on the Quantum Processors with Different Architectures}

\begin{figure}[t]
\begin{center}
\includegraphics[width=1\linewidth]{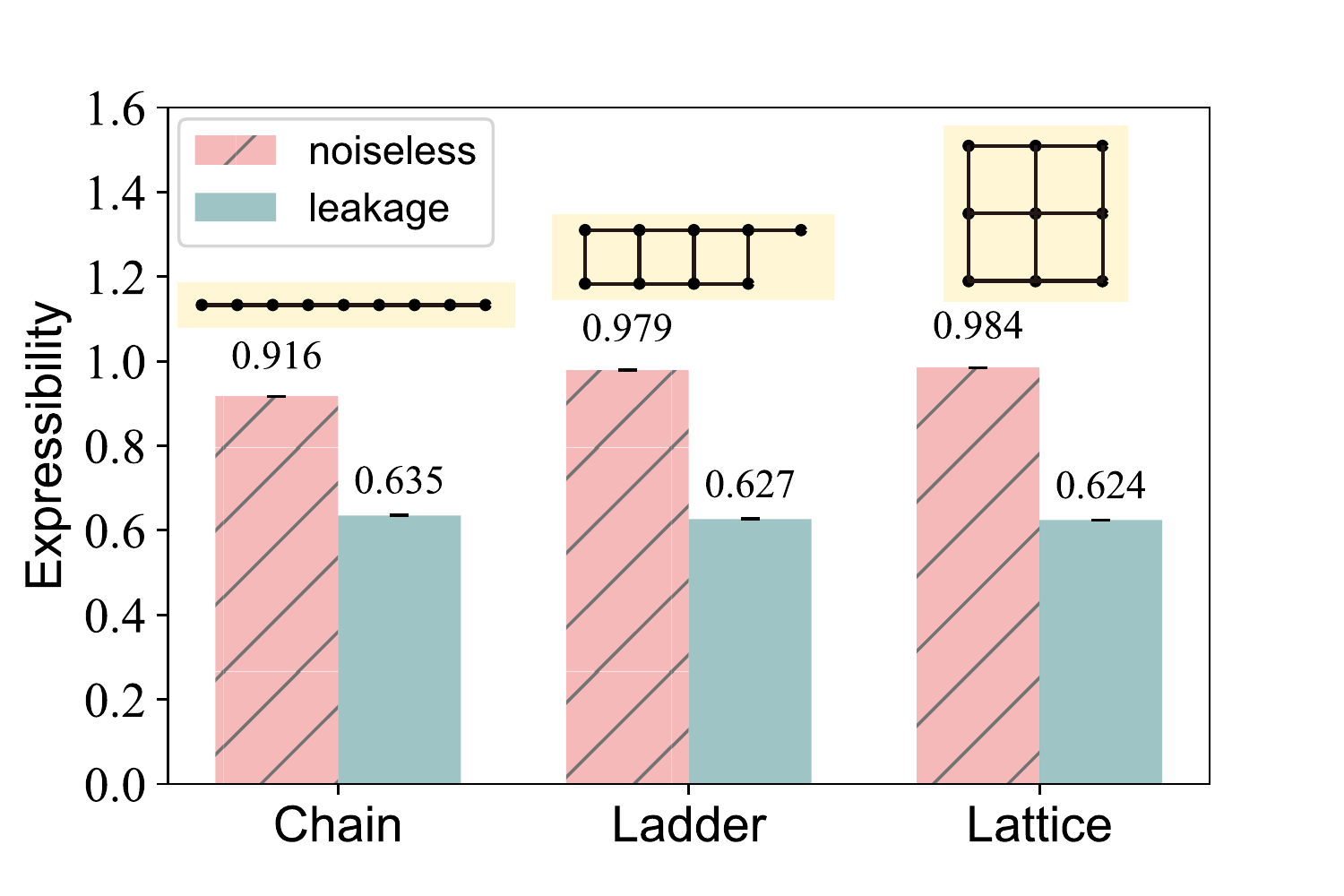}
\end{center}
\caption{\textbf{The expressibility of ansatz with chain, ladder and lattice entangling layers.} Each bar is an average over 20 independent experiments. The nearly {invisible} error bars on them suggest the numerical stability of the calculated results. The sketches above the bars show the detailed entangling layer structures, where we use CZ gates instead of CNOT.}
\label{fig:competition}
\end{figure}

Usually, the entangling layers would be slightly changed according to the topological architecture of the quantum processor, such as chain, ladder, and lattice (shown in Fig.~\ref{fig:competition}), for the purpose of improving the efficiency of the quantum hardware. Here, we assume that the an entangling layer is realized by implementing CZ gates for all nearest-neighbor coupled qubit pairs, and then compare the expressibility of the ansatz implemented on the quantum processor with various architectures. Figure~\ref{fig:competition} shows the expressibility estimation results for the ansatz with noiseless environment and leakage error, where the ansatz consists of 9 qubits and 5 depths, and the entangling layers are assumed to be implemented on the chain, ladder, and lattice processor architectures. We find that in the noiseless case, the expressibility of these ansatz rises with increasing processor connectivity, suggesting the advantage of better connectivity. However, when leakage error occurs, such advantage vanishes, as the expressibilities of the three ansatz drop to the same level.

\section{Conclusion}
To evaluate the resillience of VQA to leakage noise, we quantify the ansatz expressibility and perform comparative tests in different types of machine learning tasks. The results clearly show the detrimental effect of leakage error on VQA, as it decreases the expressibility and lowers the training accuracy. {In the numerical experiments, the model we choose is typical in terms of the hardware-efficient ansatz and the leakage occurrence in superconducting transmons~\cite{varbanov_leakage_2020,PhysRevA.97.032306,PhysRevLett.103.110501,PhysRevLett.116.020501}. Furthermore, we note that the calculation of measurement outcomes in Sec.~\ref{subsec:leakage_error_in_VQA} is independent of specific leakage occurrence models, and the expressibility evaluation in \ref{subsec:expressibility_est} is independent of the choice of cost functions and training process of VQA. Thereby, our results are not only typical, but also has a certain generality.} The rising of performance degradation with the increasing of circuit depth and leakage probability suggests the prominence of developing the leakage-reduction scheme for qubits and quantum operations, both for near- and long-term practical quantum computing.

Besides, the generalized expressibility measure we proposed is {designed from the} perspective of quantum machine learning with real-valued outputs, and can in principle be adapted to analyze a wide range of various VQA with different architectures.

The case considered in our work is the VQAs based a two-level quantum system in $\ket{0}$ and $\ket{1}$ space. The role of leakage error in high-dimensional VQAs, which employ high-dimensional quantum manipulation and detection techniques, may be substantially different, and it may even constitute a physical activation function. Thus, the impact of leakage in different scenarios still needs to be re-characterized.

\begin{acknowledgments}
H.-L. H. acknowledges support from the Youth Talent Lifting Project (Grant No. 2020-JCJQ-QT-030), National Natural Science Foundation of China (Grants No. 11905294, 12274464), China Postdoctoral Science Foundation, and the Open Research Fund from State Key Laboratory of High Performance Computing of China (Grant No. 201901-01).
\end{acknowledgments}

\bibliographystyle{apsrev4-2}
\bibliography{b}

\begin{thebibliography}{67}%
\makeatletter
\providecommand \@ifxundefined [1]{%
 \@ifx{#1\undefined}
}%
\providecommand \@ifnum [1]{%
 \ifnum #1\expandafter \@firstoftwo
 \else \expandafter \@secondoftwo
 \fi
}%
\providecommand \@ifx [1]{%
 \ifx #1\expandafter \@firstoftwo
 \else \expandafter \@secondoftwo
 \fi
}%
\providecommand \natexlab [1]{#1}%
\providecommand \enquote  [1]{``#1''}%
\providecommand \bibnamefont  [1]{#1}%
\providecommand \bibfnamefont [1]{#1}%
\providecommand \citenamefont [1]{#1}%
\providecommand \href@noop [0]{\@secondoftwo}%
\providecommand \href [0]{\begingroup \@sanitize@url \@href}%
\providecommand \@href[1]{\@@startlink{#1}\@@href}%
\providecommand \@@href[1]{\endgroup#1\@@endlink}%
\providecommand \@sanitize@url [0]{\catcode `\\12\catcode `\$12\catcode
  `\&12\catcode `\#12\catcode `\^12\catcode `\_12\catcode `\%12\relax}%
\providecommand \@@startlink[1]{}%
\providecommand \@@endlink[0]{}%
\providecommand \url  [0]{\begingroup\@sanitize@url \@url }%
\providecommand \@url [1]{\endgroup\@href {#1}{\urlprefix }}%
\providecommand \urlprefix  [0]{URL }%
\providecommand \Eprint [0]{\href }%
\providecommand \doibase [0]{https://doi.org/}%
\providecommand \selectlanguage [0]{\@gobble}%
\providecommand \bibinfo  [0]{\@secondoftwo}%
\providecommand \bibfield  [0]{\@secondoftwo}%
\providecommand \translation [1]{[#1]}%
\providecommand \BibitemOpen [0]{}%
\providecommand \bibitemStop [0]{}%
\providecommand \bibitemNoStop [0]{.\EOS\space}%
\providecommand \EOS [0]{\spacefactor3000\relax}%
\providecommand \BibitemShut  [1]{\csname bibitem#1\endcsname}%
\let\auto@bib@innerbib\@empty
\bibitem [{\citenamefont {Huang}\ \emph {et~al.}(2020)\citenamefont {Huang},
  \citenamefont {Wu}, \citenamefont {Fan},\ and\ \citenamefont
  {Zhu}}]{superconducting_2020review}%
  \BibitemOpen
  \bibfield  {author} {\bibinfo {author} {\bibfnamefont {H.-L.}\ \bibnamefont
  {Huang}}, \bibinfo {author} {\bibfnamefont {D.}~\bibnamefont {Wu}}, \bibinfo
  {author} {\bibfnamefont {D.}~\bibnamefont {Fan}},\ and\ \bibinfo {author}
  {\bibfnamefont {X.}~\bibnamefont {Zhu}},\ }\href
  {https://doi.org/10.1007/s11432-020-2881-9} {\bibfield  {journal} {\bibinfo
  {journal} {Sci. China Inf. Sci.}\ }\textbf {\bibinfo {volume} {63}},\
  \bibinfo {pages} {180501} (\bibinfo {year} {2020})}\BibitemShut {NoStop}%
\bibitem [{\citenamefont {Arute}\ \emph {et~al.}(2019)\citenamefont {Arute},
  \citenamefont {Arya}, \citenamefont {Babbush}, \citenamefont {Bacon},
  \citenamefont {Bardin}, \citenamefont {Barends}, \citenamefont {Biswas},
  \citenamefont {Boixo}, \citenamefont {Brandao}, \citenamefont {Buell} \emph
  {et~al.}}]{google_supremacy}%
  \BibitemOpen
  \bibfield  {author} {\bibinfo {author} {\bibfnamefont {F.}~\bibnamefont
  {Arute}}, \bibinfo {author} {\bibfnamefont {K.}~\bibnamefont {Arya}},
  \bibinfo {author} {\bibfnamefont {R.}~\bibnamefont {Babbush}}, \bibinfo
  {author} {\bibfnamefont {D.}~\bibnamefont {Bacon}}, \bibinfo {author}
  {\bibfnamefont {J.~C.}\ \bibnamefont {Bardin}}, \bibinfo {author}
  {\bibfnamefont {R.}~\bibnamefont {Barends}}, \bibinfo {author} {\bibfnamefont
  {R.}~\bibnamefont {Biswas}}, \bibinfo {author} {\bibfnamefont
  {S.}~\bibnamefont {Boixo}}, \bibinfo {author} {\bibfnamefont {F.~G. S.~L.}\
  \bibnamefont {Brandao}}, \bibinfo {author} {\bibfnamefont {D.~A.}\
  \bibnamefont {Buell}}, \emph {et~al.},\ }\href
  {https://doi.org/10.1038/s41586-019-1666-5} {\bibfield  {journal} {\bibinfo
  {journal} {Nature}\ }\textbf {\bibinfo {volume} {574}},\ \bibinfo {pages}
  {505} (\bibinfo {year} {2019})}\BibitemShut {NoStop}%
\bibitem [{\citenamefont {Zhong}\ \emph {et~al.}(2020)\citenamefont {Zhong},
  \citenamefont {Wang}, \citenamefont {Deng}, \citenamefont {Chen},
  \citenamefont {Peng}, \citenamefont {Luo}, \citenamefont {Qin}, \citenamefont
  {Wu}, \citenamefont {Ding}, \citenamefont {Hu} \emph
  {et~al.}}]{ustc_supremacy}%
  \BibitemOpen
  \bibfield  {author} {\bibinfo {author} {\bibfnamefont {H.-S.}\ \bibnamefont
  {Zhong}}, \bibinfo {author} {\bibfnamefont {H.}~\bibnamefont {Wang}},
  \bibinfo {author} {\bibfnamefont {Y.-H.}\ \bibnamefont {Deng}}, \bibinfo
  {author} {\bibfnamefont {M.-C.}\ \bibnamefont {Chen}}, \bibinfo {author}
  {\bibfnamefont {L.-C.}\ \bibnamefont {Peng}}, \bibinfo {author}
  {\bibfnamefont {Y.-H.}\ \bibnamefont {Luo}}, \bibinfo {author} {\bibfnamefont
  {J.}~\bibnamefont {Qin}}, \bibinfo {author} {\bibfnamefont {D.}~\bibnamefont
  {Wu}}, \bibinfo {author} {\bibfnamefont {X.}~\bibnamefont {Ding}}, \bibinfo
  {author} {\bibfnamefont {Y.}~\bibnamefont {Hu}}, \emph {et~al.},\ }\href
  {https://doi.org/10.1126/science.abe8770} {\bibfield  {journal} {\bibinfo
  {journal} {Science}\ }\textbf {\bibinfo {volume} {370}},\ \bibinfo {pages}
  {1460} (\bibinfo {year} {2020})}\BibitemShut {NoStop}%
\bibitem [{\citenamefont {Wu}\ \emph {et~al.}(2021)\citenamefont {Wu},
  \citenamefont {Bao}, \citenamefont {Cao}, \citenamefont {Chen}, \citenamefont
  {Chen}, \citenamefont {Chen}, \citenamefont {Chung}, \citenamefont {Deng},
  \citenamefont {Du}, \citenamefont {Fan} \emph {et~al.}}]{wu_strong_2021}%
  \BibitemOpen
  \bibfield  {author} {\bibinfo {author} {\bibfnamefont {Y.}~\bibnamefont
  {Wu}}, \bibinfo {author} {\bibfnamefont {W.-S.}\ \bibnamefont {Bao}},
  \bibinfo {author} {\bibfnamefont {S.}~\bibnamefont {Cao}}, \bibinfo {author}
  {\bibfnamefont {F.}~\bibnamefont {Chen}}, \bibinfo {author} {\bibfnamefont
  {M.-C.}\ \bibnamefont {Chen}}, \bibinfo {author} {\bibfnamefont
  {X.}~\bibnamefont {Chen}}, \bibinfo {author} {\bibfnamefont {T.-H.}\
  \bibnamefont {Chung}}, \bibinfo {author} {\bibfnamefont {H.}~\bibnamefont
  {Deng}}, \bibinfo {author} {\bibfnamefont {Y.}~\bibnamefont {Du}}, \bibinfo
  {author} {\bibfnamefont {D.}~\bibnamefont {Fan}}, \emph {et~al.},\ }\href
  {https://doi.org/10.1103/PhysRevLett.127.180501} {\bibfield  {journal}
  {\bibinfo  {journal} {Phys. Rev. Lett.}\ }\textbf {\bibinfo {volume} {127}},\
  \bibinfo {pages} {180501} (\bibinfo {year} {2021})}\BibitemShut {NoStop}%
\bibitem [{\citenamefont {Zhu}\ \emph {et~al.}(2022)\citenamefont {Zhu},
  \citenamefont {Cao}, \citenamefont {Chen}, \citenamefont {Chen},
  \citenamefont {Chen}, \citenamefont {Chung}, \citenamefont {Deng},
  \citenamefont {Du}, \citenamefont {Fan}, \citenamefont {Gong} \emph
  {et~al.}}]{zhu2022quantum}%
  \BibitemOpen
  \bibfield  {author} {\bibinfo {author} {\bibfnamefont {Q.}~\bibnamefont
  {Zhu}}, \bibinfo {author} {\bibfnamefont {S.}~\bibnamefont {Cao}}, \bibinfo
  {author} {\bibfnamefont {F.}~\bibnamefont {Chen}}, \bibinfo {author}
  {\bibfnamefont {M.-C.}\ \bibnamefont {Chen}}, \bibinfo {author}
  {\bibfnamefont {X.}~\bibnamefont {Chen}}, \bibinfo {author} {\bibfnamefont
  {T.-H.}\ \bibnamefont {Chung}}, \bibinfo {author} {\bibfnamefont
  {H.}~\bibnamefont {Deng}}, \bibinfo {author} {\bibfnamefont {Y.}~\bibnamefont
  {Du}}, \bibinfo {author} {\bibfnamefont {D.}~\bibnamefont {Fan}}, \bibinfo
  {author} {\bibfnamefont {M.}~\bibnamefont {Gong}}, \emph {et~al.},\ }\href
  {https://www.sciencedirect.com/science/article/pii/S2095927321006733}
  {\bibfield  {journal} {\bibinfo  {journal} {Sci. Bull.}\ }\textbf {\bibinfo
  {volume} {67}},\ \bibinfo {pages} {240} (\bibinfo {year} {2022})}\BibitemShut
  {NoStop}%
\bibitem [{\citenamefont {Guo}\ \emph {et~al.}(2021)\citenamefont {Guo},
  \citenamefont {Zhao},\ and\ \citenamefont {Huang}}]{Verifying_RQC}%
  \BibitemOpen
  \bibfield  {author} {\bibinfo {author} {\bibfnamefont {C.}~\bibnamefont
  {Guo}}, \bibinfo {author} {\bibfnamefont {Y.}~\bibnamefont {Zhao}},\ and\
  \bibinfo {author} {\bibfnamefont {H.-L.}\ \bibnamefont {Huang}},\ }\href
  {https://doi.org/10.1103/PhysRevLett.126.070502} {\bibfield  {journal}
  {\bibinfo  {journal} {Phys. Rev. Lett.}\ }\textbf {\bibinfo {volume} {126}},\
  \bibinfo {pages} {070502} (\bibinfo {year} {2021})}\BibitemShut {NoStop}%
\bibitem [{\citenamefont {Fowler}\ \emph {et~al.}(2012)\citenamefont {Fowler},
  \citenamefont {Mariantoni}, \citenamefont {Martinis},\ and\ \citenamefont
  {Cleland}}]{fowler2012surface}%
  \BibitemOpen
  \bibfield  {author} {\bibinfo {author} {\bibfnamefont {A.~G.}\ \bibnamefont
  {Fowler}}, \bibinfo {author} {\bibfnamefont {M.}~\bibnamefont {Mariantoni}},
  \bibinfo {author} {\bibfnamefont {J.~M.}\ \bibnamefont {Martinis}},\ and\
  \bibinfo {author} {\bibfnamefont {A.~N.}\ \bibnamefont {Cleland}},\ }\href
  {https://journals.aps.org/pra/abstract/10.1103/PhysRevA.86.032324} {\bibfield
   {journal} {\bibinfo  {journal} {Phys. Rev. A}\ }\textbf {\bibinfo {volume}
  {86}},\ \bibinfo {pages} {032324} (\bibinfo {year} {2012})}\BibitemShut
  {NoStop}%
\bibitem [{\citenamefont {AI}(2021)}]{chen_exponential_2021}%
  \BibitemOpen
  \bibfield  {author} {\bibinfo {author} {\bibfnamefont {G.~Q.}\ \bibnamefont
  {AI}},\ }\href {https://doi.org/10.1038/s41586-021-03588-y} {\bibfield
  {journal} {\bibinfo  {journal} {Nature}\ }\textbf {\bibinfo {volume} {595}},\
  \bibinfo {pages} {383} (\bibinfo {year} {2021})}\BibitemShut {NoStop}%
\bibitem [{\citenamefont {Zhao}\ \emph {et~al.}(2021)\citenamefont {Zhao},
  \citenamefont {Ye}, \citenamefont {Huang}, \citenamefont {Zhang},
  \citenamefont {Wu}, \citenamefont {Guan}, \citenamefont {Zhu}, \citenamefont
  {Wei}, \citenamefont {He}, \citenamefont {Cao} \emph
  {et~al.}}]{zhao_realizing_2021}%
  \BibitemOpen
  \bibfield  {author} {\bibinfo {author} {\bibfnamefont {Y.}~\bibnamefont
  {Zhao}}, \bibinfo {author} {\bibfnamefont {Y.}~\bibnamefont {Ye}}, \bibinfo
  {author} {\bibfnamefont {H.-L.}\ \bibnamefont {Huang}}, \bibinfo {author}
  {\bibfnamefont {Y.}~\bibnamefont {Zhang}}, \bibinfo {author} {\bibfnamefont
  {D.}~\bibnamefont {Wu}}, \bibinfo {author} {\bibfnamefont {H.}~\bibnamefont
  {Guan}}, \bibinfo {author} {\bibfnamefont {Q.}~\bibnamefont {Zhu}}, \bibinfo
  {author} {\bibfnamefont {Z.}~\bibnamefont {Wei}}, \bibinfo {author}
  {\bibfnamefont {T.}~\bibnamefont {He}}, \bibinfo {author} {\bibfnamefont
  {S.}~\bibnamefont {Cao}}, \emph {et~al.},\ }\href
  {http://arxiv.org/abs/2112.13505} {\bibfield  {journal} {\bibinfo  {journal}
  {arXiv:2112.13505}\ } (\bibinfo {year} {2021})}\BibitemShut {NoStop}%
\bibitem [{\citenamefont {Erhard}\ \emph {et~al.}(2021)\citenamefont {Erhard},
  \citenamefont {Nautrup}, \citenamefont {Meth}, \citenamefont {Postler},
  \citenamefont {Stricker}, \citenamefont {Stadler}, \citenamefont
  {Negnevitsky}, \citenamefont {Ringbauer}, \citenamefont {Schindler},
  \citenamefont {Briegel} \emph {et~al.}}]{erhard2021entangling}%
  \BibitemOpen
  \bibfield  {author} {\bibinfo {author} {\bibfnamefont {A.}~\bibnamefont
  {Erhard}}, \bibinfo {author} {\bibfnamefont {H.~P.}\ \bibnamefont {Nautrup}},
  \bibinfo {author} {\bibfnamefont {M.}~\bibnamefont {Meth}}, \bibinfo {author}
  {\bibfnamefont {L.}~\bibnamefont {Postler}}, \bibinfo {author} {\bibfnamefont
  {R.}~\bibnamefont {Stricker}}, \bibinfo {author} {\bibfnamefont
  {M.}~\bibnamefont {Stadler}}, \bibinfo {author} {\bibfnamefont
  {V.}~\bibnamefont {Negnevitsky}}, \bibinfo {author} {\bibfnamefont
  {M.}~\bibnamefont {Ringbauer}}, \bibinfo {author} {\bibfnamefont
  {P.}~\bibnamefont {Schindler}}, \bibinfo {author} {\bibfnamefont {H.~J.}\
  \bibnamefont {Briegel}}, \emph {et~al.},\ }\href
  {https://www.nature.com/articles/s41586-020-03079-6} {\bibfield  {journal}
  {\bibinfo  {journal} {Nature}\ }\textbf {\bibinfo {volume} {589}},\ \bibinfo
  {pages} {220} (\bibinfo {year} {2021})}\BibitemShut {NoStop}%
\bibitem [{\citenamefont {Andersen}\ \emph {et~al.}(2020)\citenamefont
  {Andersen}, \citenamefont {Remm}, \citenamefont {Lazar}, \citenamefont
  {Krinner}, \citenamefont {Lacroix}, \citenamefont {Norris}, \citenamefont
  {Gabureac}, \citenamefont {Eichler},\ and\ \citenamefont
  {Wallraff}}]{andersen_repeated_2020}%
  \BibitemOpen
  \bibfield  {author} {\bibinfo {author} {\bibfnamefont {C.~K.}\ \bibnamefont
  {Andersen}}, \bibinfo {author} {\bibfnamefont {A.}~\bibnamefont {Remm}},
  \bibinfo {author} {\bibfnamefont {S.}~\bibnamefont {Lazar}}, \bibinfo
  {author} {\bibfnamefont {S.}~\bibnamefont {Krinner}}, \bibinfo {author}
  {\bibfnamefont {N.}~\bibnamefont {Lacroix}}, \bibinfo {author} {\bibfnamefont
  {G.~J.}\ \bibnamefont {Norris}}, \bibinfo {author} {\bibfnamefont
  {M.}~\bibnamefont {Gabureac}}, \bibinfo {author} {\bibfnamefont
  {C.}~\bibnamefont {Eichler}},\ and\ \bibinfo {author} {\bibfnamefont
  {A.}~\bibnamefont {Wallraff}},\ }\href
  {https://doi.org/10.1038/s41567-020-0920-y} {\bibfield  {journal} {\bibinfo
  {journal} {Nat. Phys.}\ }\textbf {\bibinfo {volume} {16}},\ \bibinfo {pages}
  {875} (\bibinfo {year} {2020})}\BibitemShut {NoStop}%
\bibitem [{\citenamefont {Marques}\ \emph {et~al.}(2022)\citenamefont
  {Marques}, \citenamefont {Varbanov}, \citenamefont {Moreira}, \citenamefont
  {Ali}, \citenamefont {Muthusubramanian}, \citenamefont {Zachariadis},
  \citenamefont {Battistel}, \citenamefont {Beekman}, \citenamefont {Haider},
  \citenamefont {Vlothuizen} \emph {et~al.}}]{marques2021logical}%
  \BibitemOpen
  \bibfield  {author} {\bibinfo {author} {\bibfnamefont {J.}~\bibnamefont
  {Marques}}, \bibinfo {author} {\bibfnamefont {B.}~\bibnamefont {Varbanov}},
  \bibinfo {author} {\bibfnamefont {M.}~\bibnamefont {Moreira}}, \bibinfo
  {author} {\bibfnamefont {H.}~\bibnamefont {Ali}}, \bibinfo {author}
  {\bibfnamefont {N.}~\bibnamefont {Muthusubramanian}}, \bibinfo {author}
  {\bibfnamefont {C.}~\bibnamefont {Zachariadis}}, \bibinfo {author}
  {\bibfnamefont {F.}~\bibnamefont {Battistel}}, \bibinfo {author}
  {\bibfnamefont {M.}~\bibnamefont {Beekman}}, \bibinfo {author} {\bibfnamefont
  {N.}~\bibnamefont {Haider}}, \bibinfo {author} {\bibfnamefont
  {W.}~\bibnamefont {Vlothuizen}}, \emph {et~al.},\ }\href
  {https://www.nature.com/articles/s41567-021-01423-9} {\bibfield  {journal}
  {\bibinfo  {journal} {Nat. Phys.}\ }\textbf {\bibinfo {volume} {18}},\
  \bibinfo {pages} {80} (\bibinfo {year} {2022})}\BibitemShut {NoStop}%
\bibitem [{\citenamefont {Krinner}\ \emph {et~al.}(2021)\citenamefont
  {Krinner}, \citenamefont {Lacroix}, \citenamefont {Remm}, \citenamefont
  {Di~Paolo}, \citenamefont {Genois}, \citenamefont {Leroux}, \citenamefont
  {Hellings}, \citenamefont {Lazar}, \citenamefont {Swiadek}, \citenamefont
  {Herrmann} \emph {et~al.}}]{krinner2021realizing}%
  \BibitemOpen
  \bibfield  {author} {\bibinfo {author} {\bibfnamefont {S.}~\bibnamefont
  {Krinner}}, \bibinfo {author} {\bibfnamefont {N.}~\bibnamefont {Lacroix}},
  \bibinfo {author} {\bibfnamefont {A.}~\bibnamefont {Remm}}, \bibinfo {author}
  {\bibfnamefont {A.}~\bibnamefont {Di~Paolo}}, \bibinfo {author}
  {\bibfnamefont {E.}~\bibnamefont {Genois}}, \bibinfo {author} {\bibfnamefont
  {C.}~\bibnamefont {Leroux}}, \bibinfo {author} {\bibfnamefont
  {C.}~\bibnamefont {Hellings}}, \bibinfo {author} {\bibfnamefont
  {S.}~\bibnamefont {Lazar}}, \bibinfo {author} {\bibfnamefont
  {F.}~\bibnamefont {Swiadek}}, \bibinfo {author} {\bibfnamefont
  {J.}~\bibnamefont {Herrmann}}, \emph {et~al.},\ }\href
  {https://arxiv.org/abs/2112.03708} {\bibfield  {journal} {\bibinfo  {journal}
  {arXiv:2112.03708}\ } (\bibinfo {year} {2021})}\BibitemShut {NoStop}%
\bibitem [{\citenamefont {Geller}(2021)}]{PhysRevLett.127.090502}%
  \BibitemOpen
  \bibfield  {author} {\bibinfo {author} {\bibfnamefont {M.~R.}\ \bibnamefont
  {Geller}},\ }\href {https://doi.org/10.1103/PhysRevLett.127.090502}
  {\bibfield  {journal} {\bibinfo  {journal} {Phys. Rev. Lett.}\ }\textbf
  {\bibinfo {volume} {127}},\ \bibinfo {pages} {090502} (\bibinfo {year}
  {2021})}\BibitemShut {NoStop}%
\bibitem [{\citenamefont {Liu}\ \emph {et~al.}(2019)\citenamefont {Liu},
  \citenamefont {Huang}, \citenamefont {Chen}, \citenamefont {Wang},
  \citenamefont {Wang}, \citenamefont {Yang}, \citenamefont {Li}, \citenamefont
  {Liu}, \citenamefont {Dowling}, \citenamefont {Byrnes} \emph
  {et~al.}}]{9code}%
  \BibitemOpen
  \bibfield  {author} {\bibinfo {author} {\bibfnamefont {C.}~\bibnamefont
  {Liu}}, \bibinfo {author} {\bibfnamefont {H.-L.}\ \bibnamefont {Huang}},
  \bibinfo {author} {\bibfnamefont {C.}~\bibnamefont {Chen}}, \bibinfo {author}
  {\bibfnamefont {B.-Y.}\ \bibnamefont {Wang}}, \bibinfo {author}
  {\bibfnamefont {X.-L.}\ \bibnamefont {Wang}}, \bibinfo {author}
  {\bibfnamefont {T.}~\bibnamefont {Yang}}, \bibinfo {author} {\bibfnamefont
  {L.}~\bibnamefont {Li}}, \bibinfo {author} {\bibfnamefont {N.-L.}\
  \bibnamefont {Liu}}, \bibinfo {author} {\bibfnamefont {J.~P.}\ \bibnamefont
  {Dowling}}, \bibinfo {author} {\bibfnamefont {T.}~\bibnamefont {Byrnes}},
  \emph {et~al.},\ }\href {https://doi.org/10.1364/OPTICA.6.000264} {\bibfield
  {journal} {\bibinfo  {journal} {Optica}\ }\textbf {\bibinfo {volume} {6}},\
  \bibinfo {pages} {264} (\bibinfo {year} {2019})}\BibitemShut {NoStop}%
\bibitem [{\citenamefont {Huang}\ \emph
  {et~al.}(2021{\natexlab{a}})\citenamefont {Huang}, \citenamefont
  {Naro\ifmmode~\dot{z}\else \.{z}\fi{}niak}, \citenamefont {Liang},
  \citenamefont {Zhao}, \citenamefont {Castellano}, \citenamefont {Gong},
  \citenamefont {Wu}, \citenamefont {Wang}, \citenamefont {Lin}, \citenamefont
  {Xu} \emph {et~al.}}]{huang2021emulating}%
  \BibitemOpen
  \bibfield  {author} {\bibinfo {author} {\bibfnamefont {H.-L.}\ \bibnamefont
  {Huang}}, \bibinfo {author} {\bibfnamefont {M.}~\bibnamefont
  {Naro\ifmmode~\dot{z}\else \.{z}\fi{}niak}}, \bibinfo {author} {\bibfnamefont
  {F.}~\bibnamefont {Liang}}, \bibinfo {author} {\bibfnamefont
  {Y.}~\bibnamefont {Zhao}}, \bibinfo {author} {\bibfnamefont {A.~D.}\
  \bibnamefont {Castellano}}, \bibinfo {author} {\bibfnamefont
  {M.}~\bibnamefont {Gong}}, \bibinfo {author} {\bibfnamefont {Y.}~\bibnamefont
  {Wu}}, \bibinfo {author} {\bibfnamefont {S.}~\bibnamefont {Wang}}, \bibinfo
  {author} {\bibfnamefont {J.}~\bibnamefont {Lin}}, \bibinfo {author}
  {\bibfnamefont {Y.}~\bibnamefont {Xu}}, \emph {et~al.},\ }\href
  {https://doi.org/10.1103/PhysRevLett.126.090502} {\bibfield  {journal}
  {\bibinfo  {journal} {Phys. Rev. Lett.}\ }\textbf {\bibinfo {volume} {126}},\
  \bibinfo {pages} {090502} (\bibinfo {year} {2021}{\natexlab{a}})}\BibitemShut
  {NoStop}%
\bibitem [{\citenamefont {Nachman}\ \emph {et~al.}(2020)\citenamefont
  {Nachman}, \citenamefont {Urbanek}, \citenamefont {de~Jong},\ and\
  \citenamefont {Bauer}}]{unfolding_readout_noise}%
  \BibitemOpen
  \bibfield  {author} {\bibinfo {author} {\bibfnamefont {B.}~\bibnamefont
  {Nachman}}, \bibinfo {author} {\bibfnamefont {M.}~\bibnamefont {Urbanek}},
  \bibinfo {author} {\bibfnamefont {W.~A.}\ \bibnamefont {de~Jong}},\ and\
  \bibinfo {author} {\bibfnamefont {C.~W.}\ \bibnamefont {Bauer}},\ }\href
  {https://doi.org/10.1038/s41534-020-00309-7} {\bibfield  {journal} {\bibinfo
  {journal} {Npj Quantum Inf.}\ }\textbf {\bibinfo {volume} {6}},\ \bibinfo
  {pages} {84} (\bibinfo {year} {2020})}\BibitemShut {NoStop}%
\bibitem [{\citenamefont {Bravyi}\ \emph {et~al.}(2021)\citenamefont {Bravyi},
  \citenamefont {Sheldon}, \citenamefont {Kandala}, \citenamefont {Mckay},\
  and\ \citenamefont {Gambetta}}]{Mitigating_readout_noise}%
  \BibitemOpen
  \bibfield  {author} {\bibinfo {author} {\bibfnamefont {S.}~\bibnamefont
  {Bravyi}}, \bibinfo {author} {\bibfnamefont {S.}~\bibnamefont {Sheldon}},
  \bibinfo {author} {\bibfnamefont {A.}~\bibnamefont {Kandala}}, \bibinfo
  {author} {\bibfnamefont {D.~C.}\ \bibnamefont {Mckay}},\ and\ \bibinfo
  {author} {\bibfnamefont {J.~M.}\ \bibnamefont {Gambetta}},\ }\href
  {https://doi.org/10.1103/PhysRevA.103.042605} {\bibfield  {journal} {\bibinfo
   {journal} {Phys. Rev. A}\ }\textbf {\bibinfo {volume} {103}},\ \bibinfo
  {pages} {042605} (\bibinfo {year} {2021})}\BibitemShut {NoStop}%
\bibitem [{\citenamefont {Nation}\ \emph {et~al.}(2021)\citenamefont {Nation},
  \citenamefont {Kang}, \citenamefont {Sundaresan},\ and\ \citenamefont
  {Gambetta}}]{nation_scalable_2021}%
  \BibitemOpen
  \bibfield  {author} {\bibinfo {author} {\bibfnamefont {P.}~\bibnamefont
  {Nation}}, \bibinfo {author} {\bibfnamefont {H.}~\bibnamefont {Kang}},
  \bibinfo {author} {\bibfnamefont {N.}~\bibnamefont {Sundaresan}},\ and\
  \bibinfo {author} {\bibfnamefont {J.}~\bibnamefont {Gambetta}},\ }\href
  {https://doi.org/10.1103/PRXQuantum.2.040326} {\bibfield  {journal} {\bibinfo
   {journal} {PRX Quantum}\ }\textbf {\bibinfo {volume} {2}} (\bibinfo {year}
  {2021})}\BibitemShut {NoStop}%
\bibitem [{\citenamefont {Smith}\ \emph {et~al.}(2021)\citenamefont {Smith},
  \citenamefont {Khosla}, \citenamefont {Self},\ and\ \citenamefont
  {Kim}}]{smith_qubit_2021}%
  \BibitemOpen
  \bibfield  {author} {\bibinfo {author} {\bibfnamefont {A.}~\bibnamefont
  {Smith}}, \bibinfo {author} {\bibfnamefont {K.}~\bibnamefont {Khosla}},
  \bibinfo {author} {\bibfnamefont {C.}~\bibnamefont {Self}},\ and\ \bibinfo
  {author} {\bibfnamefont {M.}~\bibnamefont {Kim}},\ }\href
  {https://doi.org/10.1126/sciadv.abi8009} {\bibfield  {journal} {\bibinfo
  {journal} {Sci. Adv.}\ }\textbf {\bibinfo {volume} {7}} (\bibinfo {year}
  {2021})}\BibitemShut {NoStop}%
\bibitem [{\citenamefont {Ding}\ \emph {et~al.}(2021)\citenamefont {Ding},
  \citenamefont {Niu}, \citenamefont {Zhang}, \citenamefont {Bao},\ and\
  \citenamefont {Huang}}]{ding2021noiseresistant}%
  \BibitemOpen
  \bibfield  {author} {\bibinfo {author} {\bibfnamefont {C.}~\bibnamefont
  {Ding}}, \bibinfo {author} {\bibfnamefont {Y.-F.}\ \bibnamefont {Niu}},
  \bibinfo {author} {\bibfnamefont {S.}~\bibnamefont {Zhang}}, \bibinfo
  {author} {\bibfnamefont {W.-S.}\ \bibnamefont {Bao}},\ and\ \bibinfo {author}
  {\bibfnamefont {H.-L.}\ \bibnamefont {Huang}},\ }\href
  {http://arxiv.org/abs/2109.06805} {\bibfield  {journal} {\bibinfo  {journal}
  {arXiv:2109.06805}\ } (\bibinfo {year} {2021})}\BibitemShut {NoStop}%
\bibitem [{\citenamefont {Peruzzo}\ \emph {et~al.}(2014)\citenamefont
  {Peruzzo}, \citenamefont {McClean}, \citenamefont {Shadbolt}, \citenamefont
  {Yung}, \citenamefont {Zhou}, \citenamefont {Love}, \citenamefont
  {Aspuru-Guzik},\ and\ \citenamefont {O’Brien}}]{peruzzo_variational_2014}%
  \BibitemOpen
  \bibfield  {author} {\bibinfo {author} {\bibfnamefont {A.}~\bibnamefont
  {Peruzzo}}, \bibinfo {author} {\bibfnamefont {J.}~\bibnamefont {McClean}},
  \bibinfo {author} {\bibfnamefont {P.}~\bibnamefont {Shadbolt}}, \bibinfo
  {author} {\bibfnamefont {M.-H.}\ \bibnamefont {Yung}}, \bibinfo {author}
  {\bibfnamefont {X.-Q.}\ \bibnamefont {Zhou}}, \bibinfo {author}
  {\bibfnamefont {P.~J.}\ \bibnamefont {Love}}, \bibinfo {author}
  {\bibfnamefont {A.}~\bibnamefont {Aspuru-Guzik}},\ and\ \bibinfo {author}
  {\bibfnamefont {J.~L.}\ \bibnamefont {O’Brien}},\ }\href
  {https://doi.org/10.1038/ncomms5213} {\bibfield  {journal} {\bibinfo
  {journal} {Nat. Commun.}\ }\textbf {\bibinfo {volume} {5}},\ \bibinfo {pages}
  {4213} (\bibinfo {year} {2014})}\BibitemShut {NoStop}%
\bibitem [{\citenamefont {Huang}\ \emph
  {et~al.}(2021{\natexlab{b}})\citenamefont {Huang}, \citenamefont {Du},
  \citenamefont {Gong}, \citenamefont {Zhao}, \citenamefont {Wu}, \citenamefont
  {Wang}, \citenamefont {Li}, \citenamefont {Liang}, \citenamefont {Lin},
  \citenamefont {Xu} \emph {et~al.}}]{experimental_QGAN}%
  \BibitemOpen
  \bibfield  {author} {\bibinfo {author} {\bibfnamefont {H.-L.}\ \bibnamefont
  {Huang}}, \bibinfo {author} {\bibfnamefont {Y.}~\bibnamefont {Du}}, \bibinfo
  {author} {\bibfnamefont {M.}~\bibnamefont {Gong}}, \bibinfo {author}
  {\bibfnamefont {Y.}~\bibnamefont {Zhao}}, \bibinfo {author} {\bibfnamefont
  {Y.}~\bibnamefont {Wu}}, \bibinfo {author} {\bibfnamefont {C.}~\bibnamefont
  {Wang}}, \bibinfo {author} {\bibfnamefont {S.}~\bibnamefont {Li}}, \bibinfo
  {author} {\bibfnamefont {F.}~\bibnamefont {Liang}}, \bibinfo {author}
  {\bibfnamefont {J.}~\bibnamefont {Lin}}, \bibinfo {author} {\bibfnamefont
  {Y.}~\bibnamefont {Xu}}, \emph {et~al.},\ }\href
  {https://doi.org/10.1103/PhysRevApplied.16.024051} {\bibfield  {journal}
  {\bibinfo  {journal} {Phys. Rev. Applied}\ }\textbf {\bibinfo {volume}
  {16}},\ \bibinfo {pages} {024051} (\bibinfo {year}
  {2021}{\natexlab{b}})}\BibitemShut {NoStop}%
\bibitem [{\citenamefont {Liu}\ \emph {et~al.}(2021)\citenamefont {Liu},
  \citenamefont {Lim}, \citenamefont {Wood}, \citenamefont {Huang},
  \citenamefont {Guo},\ and\ \citenamefont {Huang}}]{qccnn}%
  \BibitemOpen
  \bibfield  {author} {\bibinfo {author} {\bibfnamefont {J.}~\bibnamefont
  {Liu}}, \bibinfo {author} {\bibfnamefont {K.~H.}\ \bibnamefont {Lim}},
  \bibinfo {author} {\bibfnamefont {K.~L.}\ \bibnamefont {Wood}}, \bibinfo
  {author} {\bibfnamefont {W.}~\bibnamefont {Huang}}, \bibinfo {author}
  {\bibfnamefont {C.}~\bibnamefont {Guo}},\ and\ \bibinfo {author}
  {\bibfnamefont {H.-L.}\ \bibnamefont {Huang}},\ }\href
  {https://doi.org/10.1007/s11433-021-1734-3} {\bibfield  {journal} {\bibinfo
  {journal} {Sci. China Phys. Mech. Astron.}\ }\textbf {\bibinfo {volume}
  {64}},\ \bibinfo {pages} {290311} (\bibinfo {year} {2021})}\BibitemShut
  {NoStop}%
\bibitem [{\citenamefont {Schuld}\ and\ \citenamefont
  {Killoran}(2019)}]{Schuld2019QuantumML}%
  \BibitemOpen
  \bibfield  {author} {\bibinfo {author} {\bibfnamefont {M.}~\bibnamefont
  {Schuld}}\ and\ \bibinfo {author} {\bibfnamefont {N.}~\bibnamefont
  {Killoran}},\ }\href {https://doi.org/10.1103/PhysRevLett.122.040504}
  {\bibfield  {journal} {\bibinfo  {journal} {Phys. Rev. Lett.}\ }\textbf
  {\bibinfo {volume} {122}},\ \bibinfo {pages} {040504} (\bibinfo {year}
  {2019})}\BibitemShut {NoStop}%
\bibitem [{\citenamefont {Biamonte}\ \emph {et~al.}(2017)\citenamefont
  {Biamonte}, \citenamefont {Wittek}, \citenamefont {Pancotti}, \citenamefont
  {Rebentrost}, \citenamefont {Wiebe},\ and\ \citenamefont
  {Lloyd}}]{biamonte2017quantum}%
  \BibitemOpen
  \bibfield  {author} {\bibinfo {author} {\bibfnamefont {J.}~\bibnamefont
  {Biamonte}}, \bibinfo {author} {\bibfnamefont {P.}~\bibnamefont {Wittek}},
  \bibinfo {author} {\bibfnamefont {N.}~\bibnamefont {Pancotti}}, \bibinfo
  {author} {\bibfnamefont {P.}~\bibnamefont {Rebentrost}}, \bibinfo {author}
  {\bibfnamefont {N.}~\bibnamefont {Wiebe}},\ and\ \bibinfo {author}
  {\bibfnamefont {S.}~\bibnamefont {Lloyd}},\ }\href
  {https://doi.org/10.1038/nature23474} {\bibfield  {journal} {\bibinfo
  {journal} {Nature}\ }\textbf {\bibinfo {volume} {549}},\ \bibinfo {pages}
  {195} (\bibinfo {year} {2017})}\BibitemShut {NoStop}%
\bibitem [{\citenamefont {Havlíček}\ \emph {et~al.}(2019)\citenamefont
  {Havlíček}, \citenamefont {Córcoles}, \citenamefont {Temme}, \citenamefont
  {Harrow}, \citenamefont {Kandala}, \citenamefont {Chow},\ and\ \citenamefont
  {Gambetta}}]{havlivcek2019supervised}%
  \BibitemOpen
  \bibfield  {author} {\bibinfo {author} {\bibfnamefont {V.}~\bibnamefont
  {Havlíček}}, \bibinfo {author} {\bibfnamefont {A.~D.}\ \bibnamefont
  {Córcoles}}, \bibinfo {author} {\bibfnamefont {K.}~\bibnamefont {Temme}},
  \bibinfo {author} {\bibfnamefont {A.~W.}\ \bibnamefont {Harrow}}, \bibinfo
  {author} {\bibfnamefont {A.}~\bibnamefont {Kandala}}, \bibinfo {author}
  {\bibfnamefont {J.~M.}\ \bibnamefont {Chow}},\ and\ \bibinfo {author}
  {\bibfnamefont {J.~M.}\ \bibnamefont {Gambetta}},\ }\href
  {https://doi.org/10.1038/s41586-019-0980-2} {\bibfield  {journal} {\bibinfo
  {journal} {Nature}\ }\textbf {\bibinfo {volume} {567}},\ \bibinfo {pages}
  {209} (\bibinfo {year} {2019})}\BibitemShut {NoStop}%
\bibitem [{\citenamefont {Saggio}\ \emph {et~al.}(2021)\citenamefont {Saggio},
  \citenamefont {Asenbeck}, \citenamefont {Hamann}, \citenamefont {Strömberg},
  \citenamefont {Schiansky}, \citenamefont {Dunjko}, \citenamefont {Friis},
  \citenamefont {Harris}, \citenamefont {Hochberg}, \citenamefont {Englund}
  \emph {et~al.}}]{saggio2021experimental}%
  \BibitemOpen
  \bibfield  {author} {\bibinfo {author} {\bibfnamefont {V.}~\bibnamefont
  {Saggio}}, \bibinfo {author} {\bibfnamefont {B.~E.}\ \bibnamefont
  {Asenbeck}}, \bibinfo {author} {\bibfnamefont {A.}~\bibnamefont {Hamann}},
  \bibinfo {author} {\bibfnamefont {T.}~\bibnamefont {Strömberg}}, \bibinfo
  {author} {\bibfnamefont {P.}~\bibnamefont {Schiansky}}, \bibinfo {author}
  {\bibfnamefont {V.}~\bibnamefont {Dunjko}}, \bibinfo {author} {\bibfnamefont
  {N.}~\bibnamefont {Friis}}, \bibinfo {author} {\bibfnamefont {N.~C.}\
  \bibnamefont {Harris}}, \bibinfo {author} {\bibfnamefont {M.}~\bibnamefont
  {Hochberg}}, \bibinfo {author} {\bibfnamefont {D.}~\bibnamefont {Englund}},
  \emph {et~al.},\ }\href {https://doi.org/10.1038/s41586-021-03242-7}
  {\bibfield  {journal} {\bibinfo  {journal} {Nature}\ }\textbf {\bibinfo
  {volume} {591}},\ \bibinfo {pages} {229} (\bibinfo {year}
  {2021})}\BibitemShut {NoStop}%
\bibitem [{\citenamefont {Cerezo}\ \emph {et~al.}(2021)\citenamefont {Cerezo},
  \citenamefont {Arrasmith}, \citenamefont {Babbush}, \citenamefont {Benjamin},
  \citenamefont {Endo}, \citenamefont {Fujii}, \citenamefont {McClean},
  \citenamefont {Mitarai}, \citenamefont {Yuan}, \citenamefont {Cincio} \emph
  {et~al.}}]{cerezo_variational_2021}%
  \BibitemOpen
  \bibfield  {author} {\bibinfo {author} {\bibfnamefont {M.}~\bibnamefont
  {Cerezo}}, \bibinfo {author} {\bibfnamefont {A.}~\bibnamefont {Arrasmith}},
  \bibinfo {author} {\bibfnamefont {R.}~\bibnamefont {Babbush}}, \bibinfo
  {author} {\bibfnamefont {S.~C.}\ \bibnamefont {Benjamin}}, \bibinfo {author}
  {\bibfnamefont {S.}~\bibnamefont {Endo}}, \bibinfo {author} {\bibfnamefont
  {K.}~\bibnamefont {Fujii}}, \bibinfo {author} {\bibfnamefont {J.~R.}\
  \bibnamefont {McClean}}, \bibinfo {author} {\bibfnamefont {K.}~\bibnamefont
  {Mitarai}}, \bibinfo {author} {\bibfnamefont {X.}~\bibnamefont {Yuan}},
  \bibinfo {author} {\bibfnamefont {L.}~\bibnamefont {Cincio}}, \emph
  {et~al.},\ }\href {https://doi.org/10.1038/s42254-021-00348-9} {\bibfield
  {journal} {\bibinfo  {journal} {Nat. Rev. Phys.}\ }\textbf {\bibinfo {volume}
  {3}},\ \bibinfo {pages} {625} (\bibinfo {year} {2021})}\BibitemShut {NoStop}%
\bibitem [{\citenamefont {Zhou}\ \emph {et~al.}(2020)\citenamefont {Zhou},
  \citenamefont {Wang}, \citenamefont {Choi}, \citenamefont {Pichler},\ and\
  \citenamefont {Lukin}}]{PhysRevX.10.021067}%
  \BibitemOpen
  \bibfield  {author} {\bibinfo {author} {\bibfnamefont {L.}~\bibnamefont
  {Zhou}}, \bibinfo {author} {\bibfnamefont {S.-T.}\ \bibnamefont {Wang}},
  \bibinfo {author} {\bibfnamefont {S.}~\bibnamefont {Choi}}, \bibinfo {author}
  {\bibfnamefont {H.}~\bibnamefont {Pichler}},\ and\ \bibinfo {author}
  {\bibfnamefont {M.~D.}\ \bibnamefont {Lukin}},\ }\href
  {https://doi.org/10.1103/PhysRevX.10.021067} {\bibfield  {journal} {\bibinfo
  {journal} {Phys. Rev. X}\ }\textbf {\bibinfo {volume} {10}},\ \bibinfo
  {pages} {021067} (\bibinfo {year} {2020})}\BibitemShut {NoStop}%
\bibitem [{\citenamefont {McClean}\ \emph {et~al.}(2016)\citenamefont
  {McClean}, \citenamefont {Romero}, \citenamefont {Babbush},\ and\
  \citenamefont {Aspuru-Guzik}}]{mcclean_theory_2016}%
  \BibitemOpen
  \bibfield  {author} {\bibinfo {author} {\bibfnamefont {J.~R.}\ \bibnamefont
  {McClean}}, \bibinfo {author} {\bibfnamefont {J.}~\bibnamefont {Romero}},
  \bibinfo {author} {\bibfnamefont {R.}~\bibnamefont {Babbush}},\ and\ \bibinfo
  {author} {\bibfnamefont {A.}~\bibnamefont {Aspuru-Guzik}},\ }\href
  {https://doi.org/10.1088/1367-2630/18/2/023023} {\bibfield  {journal}
  {\bibinfo  {journal} {New J. Phys.}\ }\textbf {\bibinfo {volume} {18}},\
  \bibinfo {pages} {023023} (\bibinfo {year} {2016})}\BibitemShut {NoStop}%
\bibitem [{\citenamefont {Gentini}\ \emph {et~al.}(2020)\citenamefont
  {Gentini}, \citenamefont {Cuccoli}, \citenamefont {Pirandola}, \citenamefont
  {Verrucchi},\ and\ \citenamefont {Banchi}}]{1gentini_noise-resilient_2020}%
  \BibitemOpen
  \bibfield  {author} {\bibinfo {author} {\bibfnamefont {L.}~\bibnamefont
  {Gentini}}, \bibinfo {author} {\bibfnamefont {A.}~\bibnamefont {Cuccoli}},
  \bibinfo {author} {\bibfnamefont {S.}~\bibnamefont {Pirandola}}, \bibinfo
  {author} {\bibfnamefont {P.}~\bibnamefont {Verrucchi}},\ and\ \bibinfo
  {author} {\bibfnamefont {L.}~\bibnamefont {Banchi}},\ }\href
  {https://doi.org/10.1103/PhysRevA.102.052414} {\bibfield  {journal} {\bibinfo
   {journal} {Phys. Rev. A}\ }\textbf {\bibinfo {volume} {102}},\ \bibinfo
  {pages} {052414} (\bibinfo {year} {2020})}\BibitemShut {NoStop}%
\bibitem [{\citenamefont {Sharma}\ \emph {et~al.}(2020)\citenamefont {Sharma},
  \citenamefont {Khatri}, \citenamefont {Cerezo},\ and\ \citenamefont
  {Coles}}]{3sharma_noise_2020}%
  \BibitemOpen
  \bibfield  {author} {\bibinfo {author} {\bibfnamefont {K.}~\bibnamefont
  {Sharma}}, \bibinfo {author} {\bibfnamefont {S.}~\bibnamefont {Khatri}},
  \bibinfo {author} {\bibfnamefont {M.}~\bibnamefont {Cerezo}},\ and\ \bibinfo
  {author} {\bibfnamefont {P.~J.}\ \bibnamefont {Coles}},\ }\href
  {https://doi.org/10.1088/1367-2630/ab784c} {\bibfield  {journal} {\bibinfo
  {journal} {New J. Phys.}\ }\textbf {\bibinfo {volume} {22}},\ \bibinfo
  {pages} {043006} (\bibinfo {year} {2020})}\BibitemShut {NoStop}%
\bibitem [{\citenamefont {Cong}\ \emph {et~al.}(2019)\citenamefont {Cong},
  \citenamefont {Choi},\ and\ \citenamefont {Lukin}}]{cong2019quantum}%
  \BibitemOpen
  \bibfield  {author} {\bibinfo {author} {\bibfnamefont {I.}~\bibnamefont
  {Cong}}, \bibinfo {author} {\bibfnamefont {S.}~\bibnamefont {Choi}},\ and\
  \bibinfo {author} {\bibfnamefont {M.~D.}\ \bibnamefont {Lukin}},\ }\href
  {https://www.nature.com/articles/s41567-019-0648-8} {\bibfield  {journal}
  {\bibinfo  {journal} {Nat. Phys.}\ }\textbf {\bibinfo {volume} {15}},\
  \bibinfo {pages} {1273} (\bibinfo {year} {2019})}\BibitemShut {NoStop}%
\bibitem [{\citenamefont {Lloyd}\ and\ \citenamefont
  {Weedbrook}(2018)}]{lloyd2018quantum}%
  \BibitemOpen
  \bibfield  {author} {\bibinfo {author} {\bibfnamefont {S.}~\bibnamefont
  {Lloyd}}\ and\ \bibinfo {author} {\bibfnamefont {C.}~\bibnamefont
  {Weedbrook}},\ }\href
  {https://journals.aps.org/prl/abstract/10.1103/PhysRevLett.121.040502}
  {\bibfield  {journal} {\bibinfo  {journal} {Phys. Rev. Lett.}\ }\textbf
  {\bibinfo {volume} {121}},\ \bibinfo {pages} {040502} (\bibinfo {year}
  {2018})}\BibitemShut {NoStop}%
\bibitem [{\citenamefont {Harrigan}\ \emph {et~al.}(2021)\citenamefont
  {Harrigan}, \citenamefont {Sung}, \citenamefont {Neeley}, \citenamefont
  {Satzinger}, \citenamefont {Arute}, \citenamefont {Arya}, \citenamefont
  {Atalaya}, \citenamefont {Bardin}, \citenamefont {Barends}, \citenamefont
  {Boixo} \emph {et~al.}}]{harrigan2021quantum}%
  \BibitemOpen
  \bibfield  {author} {\bibinfo {author} {\bibfnamefont {M.~P.}\ \bibnamefont
  {Harrigan}}, \bibinfo {author} {\bibfnamefont {K.~J.}\ \bibnamefont {Sung}},
  \bibinfo {author} {\bibfnamefont {M.}~\bibnamefont {Neeley}}, \bibinfo
  {author} {\bibfnamefont {K.~J.}\ \bibnamefont {Satzinger}}, \bibinfo {author}
  {\bibfnamefont {F.}~\bibnamefont {Arute}}, \bibinfo {author} {\bibfnamefont
  {K.}~\bibnamefont {Arya}}, \bibinfo {author} {\bibfnamefont {J.}~\bibnamefont
  {Atalaya}}, \bibinfo {author} {\bibfnamefont {J.~C.}\ \bibnamefont {Bardin}},
  \bibinfo {author} {\bibfnamefont {R.}~\bibnamefont {Barends}}, \bibinfo
  {author} {\bibfnamefont {S.}~\bibnamefont {Boixo}}, \emph {et~al.},\ }\href
  {https://www.nature.com/articles/s41567-020-01105-y} {\bibfield  {journal}
  {\bibinfo  {journal} {Nat. Phys.}\ }\textbf {\bibinfo {volume} {17}},\
  \bibinfo {pages} {332} (\bibinfo {year} {2021})}\BibitemShut {NoStop}%
\bibitem [{\citenamefont {Kandala}\ \emph {et~al.}(2017)\citenamefont
  {Kandala}, \citenamefont {Mezzacapo}, \citenamefont {Temme}, \citenamefont
  {Takita}, \citenamefont {Brink}, \citenamefont {Chow},\ and\ \citenamefont
  {Gambetta}}]{kandala_hardware-efficient_2017}%
  \BibitemOpen
  \bibfield  {author} {\bibinfo {author} {\bibfnamefont {A.}~\bibnamefont
  {Kandala}}, \bibinfo {author} {\bibfnamefont {A.}~\bibnamefont {Mezzacapo}},
  \bibinfo {author} {\bibfnamefont {K.}~\bibnamefont {Temme}}, \bibinfo
  {author} {\bibfnamefont {M.}~\bibnamefont {Takita}}, \bibinfo {author}
  {\bibfnamefont {M.}~\bibnamefont {Brink}}, \bibinfo {author} {\bibfnamefont
  {J.~M.}\ \bibnamefont {Chow}},\ and\ \bibinfo {author} {\bibfnamefont
  {J.~M.}\ \bibnamefont {Gambetta}},\ }\href
  {https://doi.org/10.1038/nature23879} {\bibfield  {journal} {\bibinfo
  {journal} {Nature}\ }\textbf {\bibinfo {volume} {549}},\ \bibinfo {pages}
  {242} (\bibinfo {year} {2017})}\BibitemShut {NoStop}%
\bibitem [{\citenamefont {Quantum}\ \emph {et~al.}(2020)\citenamefont
  {Quantum}, \citenamefont {Collaborators}, \citenamefont {Arute},
  \citenamefont {Arya}, \citenamefont {Babbush}, \citenamefont {Bacon},
  \citenamefont {Bardin}, \citenamefont {Barends}, \citenamefont {Boixo},
  \citenamefont {Broughton}, \citenamefont {Buckley} \emph
  {et~al.}}]{google2020hartree}%
  \BibitemOpen
  \bibfield  {author} {\bibinfo {author} {\bibfnamefont {G.~A.}\ \bibnamefont
  {Quantum}}, \bibinfo {author} {\bibnamefont {Collaborators}}, \bibinfo
  {author} {\bibfnamefont {F.}~\bibnamefont {Arute}}, \bibinfo {author}
  {\bibfnamefont {K.}~\bibnamefont {Arya}}, \bibinfo {author} {\bibfnamefont
  {R.}~\bibnamefont {Babbush}}, \bibinfo {author} {\bibfnamefont
  {D.}~\bibnamefont {Bacon}}, \bibinfo {author} {\bibfnamefont {J.~C.}\
  \bibnamefont {Bardin}}, \bibinfo {author} {\bibfnamefont {R.}~\bibnamefont
  {Barends}}, \bibinfo {author} {\bibfnamefont {S.}~\bibnamefont {Boixo}},
  \bibinfo {author} {\bibfnamefont {M.}~\bibnamefont {Broughton}}, \bibinfo
  {author} {\bibfnamefont {B.~B.}\ \bibnamefont {Buckley}}, \emph {et~al.},\
  }\href {https://www.science.org/doi/abs/10.1126/science.abb9811} {\bibfield
  {journal} {\bibinfo  {journal} {Science}\ }\textbf {\bibinfo {volume}
  {369}},\ \bibinfo {pages} {1084} (\bibinfo {year} {2020})}\BibitemShut
  {NoStop}%
\bibitem [{\citenamefont {Romero}\ \emph {et~al.}(2018)\citenamefont {Romero},
  \citenamefont {Babbush}, \citenamefont {McClean}, \citenamefont {Hempel},
  \citenamefont {Love},\ and\ \citenamefont {Aspuru-Guzik}}]{Romero_2018}%
  \BibitemOpen
  \bibfield  {author} {\bibinfo {author} {\bibfnamefont {J.}~\bibnamefont
  {Romero}}, \bibinfo {author} {\bibfnamefont {R.}~\bibnamefont {Babbush}},
  \bibinfo {author} {\bibfnamefont {J.~R.}\ \bibnamefont {McClean}}, \bibinfo
  {author} {\bibfnamefont {C.}~\bibnamefont {Hempel}}, \bibinfo {author}
  {\bibfnamefont {P.~J.}\ \bibnamefont {Love}},\ and\ \bibinfo {author}
  {\bibfnamefont {A.}~\bibnamefont {Aspuru-Guzik}},\ }\href
  {https://doi.org/10.1088/2058-9565/aad3e4} {\bibfield  {journal} {\bibinfo
  {journal} {Quantum Sci. Technol.}\ }\textbf {\bibinfo {volume} {4}},\
  \bibinfo {pages} {014008} (\bibinfo {year} {2018})}\BibitemShut {NoStop}%
\bibitem [{\citenamefont {Cade}\ \emph {et~al.}(2020)\citenamefont {Cade},
  \citenamefont {Mineh}, \citenamefont {Montanaro},\ and\ \citenamefont
  {Stanisic}}]{PhysRevB.102.235122}%
  \BibitemOpen
  \bibfield  {author} {\bibinfo {author} {\bibfnamefont {C.}~\bibnamefont
  {Cade}}, \bibinfo {author} {\bibfnamefont {L.}~\bibnamefont {Mineh}},
  \bibinfo {author} {\bibfnamefont {A.}~\bibnamefont {Montanaro}},\ and\
  \bibinfo {author} {\bibfnamefont {S.}~\bibnamefont {Stanisic}},\ }\href
  {https://doi.org/10.1103/PhysRevB.102.235122} {\bibfield  {journal} {\bibinfo
   {journal} {Phys. Rev. B}\ }\textbf {\bibinfo {volume} {102}},\ \bibinfo
  {pages} {235122} (\bibinfo {year} {2020})}\BibitemShut {NoStop}%
\bibitem [{\citenamefont {Hempel}\ \emph {et~al.}(2018)\citenamefont {Hempel},
  \citenamefont {Maier}, \citenamefont {Romero}, \citenamefont {McClean},
  \citenamefont {Monz}, \citenamefont {Shen}, \citenamefont {Jurcevic},
  \citenamefont {Lanyon}, \citenamefont {Love}, \citenamefont {Babbush} \emph
  {et~al.}}]{PhysRevX.8.031022}%
  \BibitemOpen
  \bibfield  {author} {\bibinfo {author} {\bibfnamefont {C.}~\bibnamefont
  {Hempel}}, \bibinfo {author} {\bibfnamefont {C.}~\bibnamefont {Maier}},
  \bibinfo {author} {\bibfnamefont {J.}~\bibnamefont {Romero}}, \bibinfo
  {author} {\bibfnamefont {J.}~\bibnamefont {McClean}}, \bibinfo {author}
  {\bibfnamefont {T.}~\bibnamefont {Monz}}, \bibinfo {author} {\bibfnamefont
  {H.}~\bibnamefont {Shen}}, \bibinfo {author} {\bibfnamefont {P.}~\bibnamefont
  {Jurcevic}}, \bibinfo {author} {\bibfnamefont {B.~P.}\ \bibnamefont
  {Lanyon}}, \bibinfo {author} {\bibfnamefont {P.}~\bibnamefont {Love}},
  \bibinfo {author} {\bibfnamefont {R.}~\bibnamefont {Babbush}}, \emph
  {et~al.},\ }\href {https://doi.org/10.1103/PhysRevX.8.031022} {\bibfield
  {journal} {\bibinfo  {journal} {Phys. Rev. X}\ }\textbf {\bibinfo {volume}
  {8}},\ \bibinfo {pages} {031022} (\bibinfo {year} {2018})}\BibitemShut
  {NoStop}%
\bibitem [{\citenamefont {Nam}\ \emph {et~al.}(2020)\citenamefont {Nam},
  \citenamefont {Chen}, \citenamefont {Pisenti}, \citenamefont {Wright},
  \citenamefont {Delaney}, \citenamefont {Maslov}, \citenamefont {Brown},
  \citenamefont {Allen}, \citenamefont {Amini}, \citenamefont {Apisdorf} \emph
  {et~al.}}]{nam_ground-state_2020}%
  \BibitemOpen
  \bibfield  {author} {\bibinfo {author} {\bibfnamefont {Y.}~\bibnamefont
  {Nam}}, \bibinfo {author} {\bibfnamefont {J.-S.}\ \bibnamefont {Chen}},
  \bibinfo {author} {\bibfnamefont {N.~C.}\ \bibnamefont {Pisenti}}, \bibinfo
  {author} {\bibfnamefont {K.}~\bibnamefont {Wright}}, \bibinfo {author}
  {\bibfnamefont {C.}~\bibnamefont {Delaney}}, \bibinfo {author} {\bibfnamefont
  {D.}~\bibnamefont {Maslov}}, \bibinfo {author} {\bibfnamefont {K.~R.}\
  \bibnamefont {Brown}}, \bibinfo {author} {\bibfnamefont {S.}~\bibnamefont
  {Allen}}, \bibinfo {author} {\bibfnamefont {J.~M.}\ \bibnamefont {Amini}},
  \bibinfo {author} {\bibfnamefont {J.}~\bibnamefont {Apisdorf}}, \emph
  {et~al.},\ }\href {https://doi.org/10.1038/s41534-020-0259-3} {\bibfield
  {journal} {\bibinfo  {journal} {npj Quantum Inf.}\ }\textbf {\bibinfo
  {volume} {6}},\ \bibinfo {pages} {33} (\bibinfo {year} {2020})}\BibitemShut
  {NoStop}%
\bibitem [{\citenamefont {Saib}\ \emph {et~al.}(2021)\citenamefont {Saib},
  \citenamefont {Wallden},\ and\ \citenamefont
  {Akhalwaya}}]{5saib_effect_2021}%
  \BibitemOpen
  \bibfield  {author} {\bibinfo {author} {\bibfnamefont {W.}~\bibnamefont
  {Saib}}, \bibinfo {author} {\bibfnamefont {P.}~\bibnamefont {Wallden}},\ and\
  \bibinfo {author} {\bibfnamefont {I.}~\bibnamefont {Akhalwaya}},\ }\href
  {http://arxiv.org/abs/2108.12388} {\bibfield  {journal} {\bibinfo  {journal}
  {arXiv:2108.12388}\ } (\bibinfo {year} {2021})}\BibitemShut {NoStop}%
\bibitem [{\citenamefont {Wright}\ \emph {et~al.}(2021)\citenamefont {Wright},
  \citenamefont {Gowrishankar}, \citenamefont {Claudino}, \citenamefont
  {Lotshaw}, \citenamefont {Nguyen}, \citenamefont {McCaskey},\ and\
  \citenamefont {Humble}}]{7wright_numerical_2021}%
  \BibitemOpen
  \bibfield  {author} {\bibinfo {author} {\bibfnamefont {J.}~\bibnamefont
  {Wright}}, \bibinfo {author} {\bibfnamefont {M.}~\bibnamefont
  {Gowrishankar}}, \bibinfo {author} {\bibfnamefont {D.}~\bibnamefont
  {Claudino}}, \bibinfo {author} {\bibfnamefont {P.~C.}\ \bibnamefont
  {Lotshaw}}, \bibinfo {author} {\bibfnamefont {T.}~\bibnamefont {Nguyen}},
  \bibinfo {author} {\bibfnamefont {A.~J.}\ \bibnamefont {McCaskey}},\ and\
  \bibinfo {author} {\bibfnamefont {T.~S.}\ \bibnamefont {Humble}},\ }\href
  {http://arxiv.org/abs/2112.15540} {\bibfield  {journal} {\bibinfo  {journal}
  {arXiv:2112.15540}\ } (\bibinfo {year} {2021})}\BibitemShut {NoStop}%
\bibitem [{\citenamefont {Gowrishankar}\ \emph {et~al.}(2021)\citenamefont
  {Gowrishankar}, \citenamefont {Wright}, \citenamefont {Claudino},
  \citenamefont {Nguyen}, \citenamefont {McCaskey},\ and\ \citenamefont
  {Humble}}]{8gowrishankar_numerical_2021}%
  \BibitemOpen
  \bibfield  {author} {\bibinfo {author} {\bibfnamefont {M.}~\bibnamefont
  {Gowrishankar}}, \bibinfo {author} {\bibfnamefont {J.}~\bibnamefont
  {Wright}}, \bibinfo {author} {\bibfnamefont {D.}~\bibnamefont {Claudino}},
  \bibinfo {author} {\bibfnamefont {T.}~\bibnamefont {Nguyen}}, \bibinfo
  {author} {\bibfnamefont {A.}~\bibnamefont {McCaskey}},\ and\ \bibinfo
  {author} {\bibfnamefont {T.~S.}\ \bibnamefont {Humble}},\ }in\ \href
  {https://doi.org/10.1109/QCE52317.2021.00032} {\emph {\bibinfo {booktitle}
  {IEEE Int. Conf. Quantum Comput. Engr. (QCE)}}}\ (\bibinfo {year} {2021})\
  pp.\ \bibinfo {pages} {155--159}\BibitemShut {NoStop}%
\bibitem [{\citenamefont {Wang}\ \emph {et~al.}(2021)\citenamefont {Wang},
  \citenamefont {Fontana}, \citenamefont {Cerezo}, \citenamefont {Sharma},
  \citenamefont {Sone}, \citenamefont {Cincio},\ and\ \citenamefont
  {Coles}}]{9wang_noise-induced_2021}%
  \BibitemOpen
  \bibfield  {author} {\bibinfo {author} {\bibfnamefont {S.}~\bibnamefont
  {Wang}}, \bibinfo {author} {\bibfnamefont {E.}~\bibnamefont {Fontana}},
  \bibinfo {author} {\bibfnamefont {M.}~\bibnamefont {Cerezo}}, \bibinfo
  {author} {\bibfnamefont {K.}~\bibnamefont {Sharma}}, \bibinfo {author}
  {\bibfnamefont {A.}~\bibnamefont {Sone}}, \bibinfo {author} {\bibfnamefont
  {L.}~\bibnamefont {Cincio}},\ and\ \bibinfo {author} {\bibfnamefont
  {P.}~\bibnamefont {Coles}},\ }\href
  {https://doi.org/10.1038/s41467-021-27045-6} {\bibfield  {journal} {\bibinfo
  {journal} {Nat. Commun.}\ }\textbf {\bibinfo {volume} {12}} (\bibinfo {year}
  {2021})}\BibitemShut {NoStop}%
\bibitem [{\citenamefont {Ito}\ \emph {et~al.}(2021)\citenamefont {Ito},
  \citenamefont {Mizukami},\ and\ \citenamefont
  {Fujii}}]{6ito_universal_nodate}%
  \BibitemOpen
  \bibfield  {author} {\bibinfo {author} {\bibfnamefont {K.}~\bibnamefont
  {Ito}}, \bibinfo {author} {\bibfnamefont {W.}~\bibnamefont {Mizukami}},\ and\
  \bibinfo {author} {\bibfnamefont {K.}~\bibnamefont {Fujii}},\ }\href
  {http://arxiv.org/abs/2106.03390} {\bibfield  {journal} {\bibinfo  {journal}
  {arXiv:2106.03390}\ } (\bibinfo {year} {2021})}\BibitemShut {NoStop}%
\bibitem [{\citenamefont {Brown}\ \emph {et~al.}(2020)\citenamefont {Brown},
  \citenamefont {Cross},\ and\ \citenamefont {Brown}}]{9259963}%
  \BibitemOpen
  \bibfield  {author} {\bibinfo {author} {\bibfnamefont {N.~C.}\ \bibnamefont
  {Brown}}, \bibinfo {author} {\bibfnamefont {A.}~\bibnamefont {Cross}},\ and\
  \bibinfo {author} {\bibfnamefont {K.~R.}\ \bibnamefont {Brown}},\ }in\ \href
  {https://doi.org/10.1109/QCE49297.2020.00043} {\emph {\bibinfo {booktitle}
  {IEEE Int. Conf. Quantum Comput. Engr. (QCE)}}}\ (\bibinfo {year} {2020})\
  pp.\ \bibinfo {pages} {286--294}\BibitemShut {NoStop}%
\bibitem [{\citenamefont {Ghosh}\ \emph {et~al.}(2013)\citenamefont {Ghosh},
  \citenamefont {Fowler}, \citenamefont {Martinis},\ and\ \citenamefont
  {Geller}}]{ghosh_understanding_2013}%
  \BibitemOpen
  \bibfield  {author} {\bibinfo {author} {\bibfnamefont {J.}~\bibnamefont
  {Ghosh}}, \bibinfo {author} {\bibfnamefont {A.~G.}\ \bibnamefont {Fowler}},
  \bibinfo {author} {\bibfnamefont {J.~M.}\ \bibnamefont {Martinis}},\ and\
  \bibinfo {author} {\bibfnamefont {M.~R.}\ \bibnamefont {Geller}},\ }\href
  {https://doi.org/10.1103/PhysRevA.88.062329} {\bibfield  {journal} {\bibinfo
  {journal} {Phys. Rev. A}\ }\textbf {\bibinfo {volume} {88}},\ \bibinfo
  {pages} {062329} (\bibinfo {year} {2013})}\BibitemShut {NoStop}%
\bibitem [{\citenamefont {Holmes}\ \emph {et~al.}(2022)\citenamefont {Holmes},
  \citenamefont {Sharma}, \citenamefont {Cerezo},\ and\ \citenamefont
  {Coles}}]{PRXQuantum.3.010313}%
  \BibitemOpen
  \bibfield  {author} {\bibinfo {author} {\bibfnamefont {Z.}~\bibnamefont
  {Holmes}}, \bibinfo {author} {\bibfnamefont {K.}~\bibnamefont {Sharma}},
  \bibinfo {author} {\bibfnamefont {M.}~\bibnamefont {Cerezo}},\ and\ \bibinfo
  {author} {\bibfnamefont {P.~J.}\ \bibnamefont {Coles}},\ }\href
  {https://doi.org/10.1103/PRXQuantum.3.010313} {\bibfield  {journal} {\bibinfo
   {journal} {PRX Quantum}\ }\textbf {\bibinfo {volume} {3}},\ \bibinfo {pages}
  {010313} (\bibinfo {year} {2022})}\BibitemShut {NoStop}%
\bibitem [{\citenamefont {Tangpanitanon}\ \emph {et~al.}(2020)\citenamefont
  {Tangpanitanon}, \citenamefont {Thanasilp}, \citenamefont {Dangniam},
  \citenamefont {Lemonde},\ and\ \citenamefont
  {Angelakis}}]{tangpanitanon_expressibility_2020}%
  \BibitemOpen
  \bibfield  {author} {\bibinfo {author} {\bibfnamefont {J.}~\bibnamefont
  {Tangpanitanon}}, \bibinfo {author} {\bibfnamefont {S.}~\bibnamefont
  {Thanasilp}}, \bibinfo {author} {\bibfnamefont {N.}~\bibnamefont {Dangniam}},
  \bibinfo {author} {\bibfnamefont {M.-A.}\ \bibnamefont {Lemonde}},\ and\
  \bibinfo {author} {\bibfnamefont {D.~G.}\ \bibnamefont {Angelakis}},\ }\href
  {https://doi.org/10.1103/PhysRevResearch.2.043364} {\bibfield  {journal}
  {\bibinfo  {journal} {Phys. Rev. Research}\ }\textbf {\bibinfo {volume}
  {2}},\ \bibinfo {pages} {043364} (\bibinfo {year} {2020})}\BibitemShut
  {NoStop}%
\bibitem [{\citenamefont {Nakaji}\ and\ \citenamefont
  {Yamamoto}(2021)}]{nakaji_expressibility_2021}%
  \BibitemOpen
  \bibfield  {author} {\bibinfo {author} {\bibfnamefont {K.}~\bibnamefont
  {Nakaji}}\ and\ \bibinfo {author} {\bibfnamefont {N.}~\bibnamefont
  {Yamamoto}},\ }\href {https://doi.org/10.22331/q-2021-04-19-434} {\bibfield
  {journal} {\bibinfo  {journal} {Quantum}\ }\textbf {\bibinfo {volume} {5}},\
  \bibinfo {pages} {434} (\bibinfo {year} {2021})},\ \bibinfo {note} {arXiv:
  2005.12537}\BibitemShut {NoStop}%
\bibitem [{\citenamefont {Sim}\ \emph {et~al.}(2019)\citenamefont {Sim},
  \citenamefont {Johnson},\ and\ \citenamefont
  {Aspuru-Guzik}}]{ssim_expressibility}%
  \BibitemOpen
  \bibfield  {author} {\bibinfo {author} {\bibfnamefont {S.}~\bibnamefont
  {Sim}}, \bibinfo {author} {\bibfnamefont {P.~D.}\ \bibnamefont {Johnson}},\
  and\ \bibinfo {author} {\bibfnamefont {A.}~\bibnamefont {Aspuru-Guzik}},\
  }\href {https://onlinelibrary.wiley.com/doi/abs/10.1002/qute.201900070}
  {\bibfield  {journal} {\bibinfo  {journal} {Adv. Quantum Technol.}\ }\textbf
  {\bibinfo {volume} {2}},\ \bibinfo {pages} {1900070} (\bibinfo {year}
  {2019})}\BibitemShut {NoStop}%
\bibitem [{\citenamefont {Kullback}\ and\ \citenamefont
  {Leibler}(1951)}]{10.1214/aoms/1177729694}%
  \BibitemOpen
  \bibfield  {author} {\bibinfo {author} {\bibfnamefont {S.}~\bibnamefont
  {Kullback}}\ and\ \bibinfo {author} {\bibfnamefont {R.~A.}\ \bibnamefont
  {Leibler}},\ }\href {https://doi.org/10.1214/aoms/1177729694} {\bibfield
  {journal} {\bibinfo  {journal} {Ann. Math. Stat.}\ }\textbf {\bibinfo
  {volume} {22}},\ \bibinfo {pages} {79 } (\bibinfo {year} {1951})}\BibitemShut
  {NoStop}%
\bibitem [{\citenamefont {Gretton}\ \emph {et~al.}(2012)\citenamefont
  {Gretton}, \citenamefont {Borgwardt}, \citenamefont {Rasch}, \citenamefont
  {Scholkopf},\ and\ \citenamefont {Smola}}]{gretton_kernel_2012}%
  \BibitemOpen
  \bibfield  {author} {\bibinfo {author} {\bibfnamefont {A.}~\bibnamefont
  {Gretton}}, \bibinfo {author} {\bibfnamefont {K.}~\bibnamefont {Borgwardt}},
  \bibinfo {author} {\bibfnamefont {M.}~\bibnamefont {Rasch}}, \bibinfo
  {author} {\bibfnamefont {B.}~\bibnamefont {Scholkopf}},\ and\ \bibinfo
  {author} {\bibfnamefont {A.}~\bibnamefont {Smola}},\ }\href
  {https://www.jmlr.org/papers/volume13/gretton12a/gretton12a.pdf} {\bibfield
  {journal} {\bibinfo  {journal} {J. Mach. Learn. Res.}\ }\textbf {\bibinfo
  {volume} {13}},\ \bibinfo {pages} {723} (\bibinfo {year} {2012})}\BibitemShut
  {NoStop}%
\bibitem [{\citenamefont {Takhanov}(2021)}]{DBLP:journals/corr/abs-2106-14277}%
  \BibitemOpen
  \bibfield  {author} {\bibinfo {author} {\bibfnamefont {R.}~\bibnamefont
  {Takhanov}},\ }\href {https://arxiv.org/abs/2106.14277} {\bibfield  {journal}
  {\bibinfo  {journal} {arXiv:2106.14277}\ } (\bibinfo {year}
  {2021})}\BibitemShut {NoStop}%
\bibitem [{\citenamefont {Zoufal}\ \emph {et~al.}(2019)\citenamefont {Zoufal},
  \citenamefont {Lucchi},\ and\ \citenamefont {Woerner}}]{zoufal_quantum_2019}%
  \BibitemOpen
  \bibfield  {author} {\bibinfo {author} {\bibfnamefont {C.}~\bibnamefont
  {Zoufal}}, \bibinfo {author} {\bibfnamefont {A.}~\bibnamefont {Lucchi}},\
  and\ \bibinfo {author} {\bibfnamefont {S.}~\bibnamefont {Woerner}},\ }\href
  {https://doi.org/10.1038/s41534-019-0223-2} {\bibfield  {journal} {\bibinfo
  {journal} {Npj Quantum Inf.}\ }\textbf {\bibinfo {volume} {5}},\ \bibinfo
  {pages} {103} (\bibinfo {year} {2019})}\BibitemShut {NoStop}%
\bibitem [{\citenamefont {Campos}\ \emph {et~al.}(2021)\citenamefont {Campos},
  \citenamefont {Nasrallah},\ and\ \citenamefont
  {Biamonte}}]{PhysRevA.103.032607}%
  \BibitemOpen
  \bibfield  {author} {\bibinfo {author} {\bibfnamefont {E.}~\bibnamefont
  {Campos}}, \bibinfo {author} {\bibfnamefont {A.}~\bibnamefont {Nasrallah}},\
  and\ \bibinfo {author} {\bibfnamefont {J.}~\bibnamefont {Biamonte}},\ }\href
  {https://doi.org/10.1103/PhysRevA.103.032607} {\bibfield  {journal} {\bibinfo
   {journal} {Phys. Rev. A}\ }\textbf {\bibinfo {volume} {103}},\ \bibinfo
  {pages} {032607} (\bibinfo {year} {2021})}\BibitemShut {NoStop}%
\bibitem [{\citenamefont {Schuld}\ \emph {et~al.}(2020)\citenamefont {Schuld},
  \citenamefont {Bocharov}, \citenamefont {Svore},\ and\ \citenamefont
  {Wiebe}}]{PhysRevA.101.032308}%
  \BibitemOpen
  \bibfield  {author} {\bibinfo {author} {\bibfnamefont {M.}~\bibnamefont
  {Schuld}}, \bibinfo {author} {\bibfnamefont {A.}~\bibnamefont {Bocharov}},
  \bibinfo {author} {\bibfnamefont {K.~M.}\ \bibnamefont {Svore}},\ and\
  \bibinfo {author} {\bibfnamefont {N.}~\bibnamefont {Wiebe}},\ }\href
  {https://doi.org/10.1103/PhysRevA.101.032308} {\bibfield  {journal} {\bibinfo
   {journal} {Phys. Rev. A}\ }\textbf {\bibinfo {volume} {101}},\ \bibinfo
  {pages} {032308} (\bibinfo {year} {2020})}\BibitemShut {NoStop}%
\bibitem [{\citenamefont {Schuld}\ \emph {et~al.}(2021)\citenamefont {Schuld},
  \citenamefont {Sweke},\ and\ \citenamefont {Meyer}}]{PhysRevA.103.032430}%
  \BibitemOpen
  \bibfield  {author} {\bibinfo {author} {\bibfnamefont {M.}~\bibnamefont
  {Schuld}}, \bibinfo {author} {\bibfnamefont {R.}~\bibnamefont {Sweke}},\ and\
  \bibinfo {author} {\bibfnamefont {J.~J.}\ \bibnamefont {Meyer}},\ }\href
  {https://doi.org/10.1103/PhysRevA.103.032430} {\bibfield  {journal} {\bibinfo
   {journal} {Phys. Rev. A}\ }\textbf {\bibinfo {volume} {103}},\ \bibinfo
  {pages} {032430} (\bibinfo {year} {2021})}\BibitemShut {NoStop}%
\bibitem [{\citenamefont {Bharti}\ \emph {et~al.}(2022)\citenamefont {Bharti},
  \citenamefont {Cervera-Lierta}, \citenamefont {Kyaw}, \citenamefont {Haug},
  \citenamefont {Alperin-Lea}, \citenamefont {Anand}, \citenamefont {Degroote},
  \citenamefont {Heimonen}, \citenamefont {Kottmann}, \citenamefont {Menke},
  \citenamefont {Mok}, \citenamefont {Sim}, \citenamefont {Kwek},\ and\
  \citenamefont {Aspuru-Guzik}}]{RevModPhys.94.015004}%
  \BibitemOpen
  \bibfield  {author} {\bibinfo {author} {\bibfnamefont {K.}~\bibnamefont
  {Bharti}}, \bibinfo {author} {\bibfnamefont {A.}~\bibnamefont
  {Cervera-Lierta}}, \bibinfo {author} {\bibfnamefont {T.~H.}\ \bibnamefont
  {Kyaw}}, \bibinfo {author} {\bibfnamefont {T.}~\bibnamefont {Haug}}, \bibinfo
  {author} {\bibfnamefont {S.}~\bibnamefont {Alperin-Lea}}, \bibinfo {author}
  {\bibfnamefont {A.}~\bibnamefont {Anand}}, \bibinfo {author} {\bibfnamefont
  {M.}~\bibnamefont {Degroote}}, \bibinfo {author} {\bibfnamefont
  {H.}~\bibnamefont {Heimonen}}, \bibinfo {author} {\bibfnamefont {J.~S.}\
  \bibnamefont {Kottmann}}, \bibinfo {author} {\bibfnamefont {T.}~\bibnamefont
  {Menke}}, \bibinfo {author} {\bibfnamefont {W.-K.}\ \bibnamefont {Mok}},
  \bibinfo {author} {\bibfnamefont {S.}~\bibnamefont {Sim}}, \bibinfo {author}
  {\bibfnamefont {L.-C.}\ \bibnamefont {Kwek}},\ and\ \bibinfo {author}
  {\bibfnamefont {A.}~\bibnamefont {Aspuru-Guzik}},\ }\href
  {https://doi.org/10.1103/RevModPhys.94.015004} {\bibfield  {journal}
  {\bibinfo  {journal} {Rev. Mod. Phys.}\ }\textbf {\bibinfo {volume} {94}},\
  \bibinfo {pages} {015004} (\bibinfo {year} {2022})}\BibitemShut {NoStop}%
\bibitem [{\citenamefont {Varbanov}\ \emph {et~al.}(2020)\citenamefont
  {Varbanov}, \citenamefont {Battistel}, \citenamefont {Tarasinski},
  \citenamefont {Ostroukh}, \citenamefont {O’Brien}, \citenamefont
  {DiCarlo},\ and\ \citenamefont {Terhal}}]{varbanov_leakage_2020}%
  \BibitemOpen
  \bibfield  {author} {\bibinfo {author} {\bibfnamefont {B.~M.}\ \bibnamefont
  {Varbanov}}, \bibinfo {author} {\bibfnamefont {F.}~\bibnamefont {Battistel}},
  \bibinfo {author} {\bibfnamefont {B.~M.}\ \bibnamefont {Tarasinski}},
  \bibinfo {author} {\bibfnamefont {V.~P.}\ \bibnamefont {Ostroukh}}, \bibinfo
  {author} {\bibfnamefont {T.~E.}\ \bibnamefont {O’Brien}}, \bibinfo {author}
  {\bibfnamefont {L.}~\bibnamefont {DiCarlo}},\ and\ \bibinfo {author}
  {\bibfnamefont {B.~M.}\ \bibnamefont {Terhal}},\ }\href
  {https://doi.org/10.1038/s41534-020-00330-w} {\bibfield  {journal} {\bibinfo
  {journal} {Npj Quantum Inf.}\ }\textbf {\bibinfo {volume} {6}},\ \bibinfo
  {pages} {102} (\bibinfo {year} {2020})}\BibitemShut {NoStop}%
\bibitem [{\citenamefont {Wood}\ and\ \citenamefont
  {Gambetta}(2018)}]{PhysRevA.97.032306}%
  \BibitemOpen
  \bibfield  {author} {\bibinfo {author} {\bibfnamefont {C.~J.}\ \bibnamefont
  {Wood}}\ and\ \bibinfo {author} {\bibfnamefont {J.~M.}\ \bibnamefont
  {Gambetta}},\ }\href {https://doi.org/10.1103/PhysRevA.97.032306} {\bibfield
  {journal} {\bibinfo  {journal} {Phys. Rev. A}\ }\textbf {\bibinfo {volume}
  {97}},\ \bibinfo {pages} {032306} (\bibinfo {year} {2018})}\BibitemShut
  {NoStop}%
\bibitem [{\citenamefont {Anderson}(1936)}]{10.2307/2394164}%
  \BibitemOpen
  \bibfield  {author} {\bibinfo {author} {\bibfnamefont {E.}~\bibnamefont
  {Anderson}},\ }\href {http://www.jstor.org/stable/2394164} {\bibfield
  {journal} {\bibinfo  {journal} {Ann. Missouri Bot.}\ }\textbf {\bibinfo
  {volume} {23}},\ \bibinfo {pages} {457} (\bibinfo {year} {1936})}\BibitemShut
  {NoStop}%
\bibitem [{\citenamefont
  {FISHER}(1936)}]{https://doi.org/10.1111/j.1469-1809.1936.tb02137.x}%
  \BibitemOpen
  \bibfield  {author} {\bibinfo {author} {\bibfnamefont {R.~A.}\ \bibnamefont
  {FISHER}},\ }\href
  {https://doi.org/https://doi.org/10.1111/j.1469-1809.1936.tb02137.x}
  {\bibfield  {journal} {\bibinfo  {journal} {Ann. Eugen.}\ }\textbf {\bibinfo
  {volume} {7}},\ \bibinfo {pages} {179} (\bibinfo {year} {1936})}\BibitemShut
  {NoStop}%
\bibitem [{\citenamefont {Motzoi}\ \emph {et~al.}(2009)\citenamefont {Motzoi},
  \citenamefont {Gambetta}, \citenamefont {Rebentrost},\ and\ \citenamefont
  {Wilhelm}}]{PhysRevLett.103.110501}%
  \BibitemOpen
  \bibfield  {author} {\bibinfo {author} {\bibfnamefont {F.}~\bibnamefont
  {Motzoi}}, \bibinfo {author} {\bibfnamefont {J.~M.}\ \bibnamefont
  {Gambetta}}, \bibinfo {author} {\bibfnamefont {P.}~\bibnamefont
  {Rebentrost}},\ and\ \bibinfo {author} {\bibfnamefont {F.~K.}\ \bibnamefont
  {Wilhelm}},\ }\href {https://doi.org/10.1103/PhysRevLett.103.110501}
  {\bibfield  {journal} {\bibinfo  {journal} {Phys. Rev. Lett.}\ }\textbf
  {\bibinfo {volume} {103}},\ \bibinfo {pages} {110501} (\bibinfo {year}
  {2009})}\BibitemShut {NoStop}%
\bibitem [{\citenamefont {Chen}\ \emph {et~al.}(2016)\citenamefont {Chen},
  \citenamefont {Kelly}, \citenamefont {Quintana}, \citenamefont {Barends},
  \citenamefont {Campbell}, \citenamefont {Chen}, \citenamefont {Chiaro},
  \citenamefont {Dunsworth}, \citenamefont {Fowler}, \citenamefont {Lucero},
  \citenamefont {Jeffrey}, \citenamefont {Megrant}, \citenamefont {Mutus},
  \citenamefont {Neeley}, \citenamefont {Neill}, \citenamefont {O'Malley},
  \citenamefont {Roushan}, \citenamefont {Sank}, \citenamefont {Vainsencher},
  \citenamefont {Wenner}, \citenamefont {White}, \citenamefont {Korotkov},\
  and\ \citenamefont {Martinis}}]{PhysRevLett.116.020501}%
  \BibitemOpen
  \bibfield  {author} {\bibinfo {author} {\bibfnamefont {Z.}~\bibnamefont
  {Chen}}, \bibinfo {author} {\bibfnamefont {J.}~\bibnamefont {Kelly}},
  \bibinfo {author} {\bibfnamefont {C.}~\bibnamefont {Quintana}}, \bibinfo
  {author} {\bibfnamefont {R.}~\bibnamefont {Barends}}, \bibinfo {author}
  {\bibfnamefont {B.}~\bibnamefont {Campbell}}, \bibinfo {author}
  {\bibfnamefont {Y.}~\bibnamefont {Chen}}, \bibinfo {author} {\bibfnamefont
  {B.}~\bibnamefont {Chiaro}}, \bibinfo {author} {\bibfnamefont
  {A.}~\bibnamefont {Dunsworth}}, \bibinfo {author} {\bibfnamefont {A.~G.}\
  \bibnamefont {Fowler}}, \bibinfo {author} {\bibfnamefont {E.}~\bibnamefont
  {Lucero}}, \bibinfo {author} {\bibfnamefont {E.}~\bibnamefont {Jeffrey}},
  \bibinfo {author} {\bibfnamefont {A.}~\bibnamefont {Megrant}}, \bibinfo
  {author} {\bibfnamefont {J.}~\bibnamefont {Mutus}}, \bibinfo {author}
  {\bibfnamefont {M.}~\bibnamefont {Neeley}}, \bibinfo {author} {\bibfnamefont
  {C.}~\bibnamefont {Neill}}, \bibinfo {author} {\bibfnamefont {P.~J.~J.}\
  \bibnamefont {O'Malley}}, \bibinfo {author} {\bibfnamefont {P.}~\bibnamefont
  {Roushan}}, \bibinfo {author} {\bibfnamefont {D.}~\bibnamefont {Sank}},
  \bibinfo {author} {\bibfnamefont {A.}~\bibnamefont {Vainsencher}}, \bibinfo
  {author} {\bibfnamefont {J.}~\bibnamefont {Wenner}}, \bibinfo {author}
  {\bibfnamefont {T.~C.}\ \bibnamefont {White}}, \bibinfo {author}
  {\bibfnamefont {A.~N.}\ \bibnamefont {Korotkov}},\ and\ \bibinfo {author}
  {\bibfnamefont {J.~M.}\ \bibnamefont {Martinis}},\ }\href
  {https://doi.org/10.1103/PhysRevLett.116.020501} {\bibfield  {journal}
  {\bibinfo  {journal} {Phys. Rev. Lett.}\ }\textbf {\bibinfo {volume} {116}},\
  \bibinfo {pages} {020501} (\bibinfo {year} {2016})}\BibitemShut {NoStop}%
\end{thebibliography}%

\end{document}